\documentclass[a4paper,11pt]{article}
\usepackage[margin=0.5in]{geometry}
\usepackage{enumerate}   
\usepackage{jcappub}
\usepackage{subcaption}
\usepackage{aas_macros}
\usepackage{multirow}

\usepackage{soul}

\usepackage{amsmath}
\usepackage{amssymb}
\usepackage{bm}
\usepackage{natbib}
\usepackage{xspace}
\usepackage{lipsum}
\usepackage[utf8]{inputenc}
\usepackage[T1]{fontenc}
\usepackage{hyperref}
\usepackage{caption}
\usepackage{verbatim}
\usepackage[dvipsnames]{xcolor}
\usepackage{adjustbox}

\newcommand{\planck}{{\sl Planck}\xspace}

\newcommand{\nmt}{{\tt NaMaster}\xspace}

\newcommand{\ihMpc}{h{\rm Mpc}^{-1}}

\newcommand{\fsb}{F$^2$B\xspace}
\newcommand{\lAs}{\log\left(10^{10}A_s\right)}
\newcommand{\oc}{\Omega_ch^2}
\newcommand{\gpcoh}{h^{-1}{\rm Gpc}}
\hypersetup{
	unicode,
	pdftitle={Cosmological constraints from the angular power spectrum and bispectrum of luminous red galaxies and CMB lensing},
	pdfcreator={pdflatex}
}
\usepackage{graphicx, color}

\newcommand{\nv}{\hat{\bf n}}
\newcommand{\absum}{\textsc{AbacusSummit}\xspace}

\title{\boldmath Cosmological constraints from the angular power spectrum and bispectrum of luminous red galaxies and CMB lensing}
  \author[a,b,c,d]{Francesco Verdiani,}
  \author[e]{L\'ea Harscouet,}
  \author[e]{Matteo Zennaro,}
  \author[e]{David Alonso,}
  \author[f,g]{Boryana Hadzhiyska}

  \affiliation[a]{Scuola Internazionale Superiore di Studi Avanzati (SISSA), Via Bonomea 265, 34136 Trieste, Italy}
    \affiliation[b]{INFN Sezione di Trieste, Via Valerio 2, 34127 Trieste, Italy}
    \affiliation[c]{Istituto Nazionale di Astrofisica, Osservatorio Astronomico di Trieste, Via Tiepolo 11, 34143 Trieste, Italy}
    \affiliation[d]{Institute for Fundamental Physics of the Universe, Via Beirut 2, 34151 Trieste, Italy}
  \affiliation[e]{Department of Physics, University of Oxford, Denys Wilkinson Building, Keble Road, Oxford OX1 3RH, United Kingdom}
  \affiliation[f]{Institute of Astronomy, Madingley Road, Cambridge, CB3 0HA, United Kingdom}
  \affiliation[g]{Kavli Institute for Cosmology Cambridge, Madingley Road, Cambridge, CB3 0HA, United Kingdom}

\emailAdd{fverdian@sissa.it}

\abstract{We study the projected clustering of photometric luminous red galaxies from the DESI Legacy Survey, combining their angular power spectrum, bispectrum, and cross-correlation with maps of the CMB lensing convergence from the \planck satellite. We employ a perturbative bias expansion in Eulerian space to describe the clustering of galaxies, modelling the power spectrum and bispectrum at one-loop and tree level, respectively. This allows us to use the power spectrum to self-consistently calibrate the perturbative bias parameters. We validate this model against an $N$-body simulation, and show that it can be used up to scales of at least $k_{\rm max}^P\simeq 0.2\,\ihMpc$ and $k_{\rm max}^B\simeq 0.08\,\ihMpc$, saturating the information recovered from the data. We obtain constraints on the amplitude of matter fluctuations $\sigma_8=0.761\pm 0.020$ and the non-relativistic matter fraction $\Omega_m=0.307\pm 0.015$, as well as the combination $S_8\equiv\sigma_8\sqrt{\Omega_m/0.3}=0.769 \pm 0.020$. Including the galaxy bispectrum leads to a $10$-$20\%$ improvement on the cosmological constraints, which are also in good agreement with previous analyses of the same data, and in mild tension with \planck at the $\sim2.5\sigma$ level. This tension is largely present in the standard two-point function dataset, and the addition of the bispectrum increases it slightly, marginally shifting $\sigma_8$ downwards and $\Omega_m$ upwards. Finally, using the bispectrum allows for a substantially more precise measurement of the bias parameters of this sample, which are in reasonable agreement with existing coevolution relations.}
\begin{document}
\maketitle
\flushbottom

\section{Introduction}\label{sec:intro}
  The projected clustering of galaxies is one of the oldest probes of the large-scale structure (LSS) in existence. Early measurements of the angular correlation function of galaxies and clusters of galaxies from optical and radio surveys were vital in establishing the concept of galaxy bias, providing evidence for the large-scale homogeneity of the Universe, and confirming the basic predictions of gravitationally-led structure formation  \cite{1984ApJ...284L...9K,1990Natur.348..705E,1991MNRAS.253P...1P,1991MNRAS.253..307P,1993MNRAS.265..145B}.

  The amount of cosmological information that can be extracted from the angular clustering of galaxies in photometric surveys is relatively limited by comparison with their spectroscopic counterparts, where information from the full three-dimensional distribution of galaxies can be exploited \cite{2112.04515,2402.13310,2411.12021,2503.09714}. Nevertheless, photometric surveys have a number of advantages, most importantly the large number density of sources that can be photometrically identified. This allows us to study their clustering over smaller scales before reaching the shot noise-dominated regime, to increase our sensitivity to important effects, such as lensing magnification, and to obtain high-precision measurements of the cross-correlation between galaxies and other projected tracers of the LSS \cite{1912.08209,2105.12108,2309.05659,sailer_cosmological_2024,2503.24385,2510.09563}. In particular, the combination of these cross-correlations with the auto-correlation of galaxies at different redshifts enables a \emph{tomographic} reconstruction of the key physical quantities that these probes are sensitive to \cite{1810.00885,1909.09102,2006.14650,2504.05384,2508.05319}. Within this general framework, \emph{lensing tomography}, the combination of galaxy clustering and their cross-correlation with measurements of the gravitational lensing effect, has now become an established and high-precision cosmological probe. Lensing tomography analyses have allowed us to reconstruct the universe's growth history at high precision, reaching redshifts $z\sim3$ \cite{2103.15862,2402.05761,2507.08798}, as well as measuring the cosmological parameters governing the distribution evolution of cosmic structures at late times, such as the matter fraction $\Omega_{\rm m}$, or the expansion rate $H_0$.

  Lensing tomography, as well as other data combinations employing the angular clustering of galaxies, are set to become more important with the advent of Stage-IV \cite{astro-ph/0609591} imaging surveys, such as Euclid \cite{1110.3193}, the Rubin Observatory \cite{0805.2366,1809.01669}, and the future Roman Space telescope \cite{1503.03757}, as well as surveys at other wavelengths. It is therefore important to ensure that as much information as possible is recovered from these observations, including any potentially new effect that could become observable given the increased sensitivity that these experiments will achieve. In particular, the study of higher-order correlators of the projected galaxy distribution, such as the bispectrum, has lagged behind the use of similar statistics in other probes, such as spectroscopic clustering \cite{1407.5668,1606.00439,2010.06179,2206.02800}, the Cosmic Microwave Background (CMB) \cite{1509.08107,1905.05697}, and cosmic shear \cite{2110.10141,2208.11686}. In part this is due to the relative complexity inherent to bispectrum analysis. First, bispectrum estimators are typically significantly more computationally costly than power spectrum methods. Secondly, the resulting data vector is often much larger, involving a large number of triangle configurations. Thirdly, estimating the covariance matrix of the measurements using simulations can be prohibitively expensive. Finally, the regime over which different models for the galaxy bispectrum are sufficiently accurate has not been as thoroughly studied as the case of the power spectrum, and obtaining theoretical predictions is also often expensive. The first three problems in this list may be tackled through the use of novel bispectrum estimators \cite{2004.03574,2202.11724,2303.08828,2311.04213}, such as the ``filter-square'' approach first presented in \cite{harscouet_fast_2025}. Likewise, advances in machine learning-based emulators may be used to accelerate the theoretical calculations.

  In this paper we carry out the first comprehensive analysis of the angular power spectrum and bispectrum of galaxies in combination with CMB lensing. Our work will build on the analysis of \cite{harscouet_constraints_2025} (H25 hereafter), where the measurements of the galaxy power- and bispectrum used here, as well as their first rudimentary analysis, were presented. Crucially, we will use simulations to quantify the regime over which a perturbative treatment of the galaxy bias may be safely used in a joint analysis of 2- and 3-point correlators, before applying this framework to the real data. We will show that including the bispectrum allows us to self-consistently measure the free parameters of the bias expansion at high precision, thereby improving the final cosmological constraints, while demonstrating the internal consistency of the model.

  This paper is structured as follows. Section \ref{sec:data} presents the datasets used, as well as the methods used to generate the power spectrum and bispectrum measurements used in our analysis. The model used to describe these measurements, the simulations used to validate this model, and the method used to obtain constraints on the model parameters are described in Section \ref{sec:meth}. The results of our analysis, including both the validation on simulations and the constraints found on real data, are presented in Section \ref{sec:res}. We summarise our main results and conclude in Section \ref{sec:conc}.

\section{Data}\label{sec:data}
  \subsection{Public galaxy and CMB lensing data}\label{ssec:data.data}
    We use the publicly available \planck PR4 CMB lensing map, specifically the Generalized Minimum Variance (GMV) map, constructed from a joint Wiener filtering of the temperature and polarisation maps. The angular multipoles of the map are first rotated to Equatorial coordinates before being low-pass filtered with a top-hat filter that nulls all harmonic coefficients with $\ell > 3N_{\rm side}$ and being transformed into a real-space map. Here $N_{\rm side}=2048$ is the {\tt HEALPix} pixelisation resolution parameter used, corresponding to a pixel size of $\delta\theta\simeq1.7'$. Similarly, the publicly available sky map associated with this map is rotated to Equatorial coordinates in pixel space, and then apodised using a ``C1'' apodisation kernel with angular scale $0.2^\circ$, as implemented in {\tt NaMaster}\footnote{\url{https://github.com/LSSTDESC/NaMaster}}. We then make the resulting mask binary, setting all values above $0.5$ to 1, and all other values to zero.

\begin{figure}[b]
  \centering
  \includegraphics[width=\linewidth]{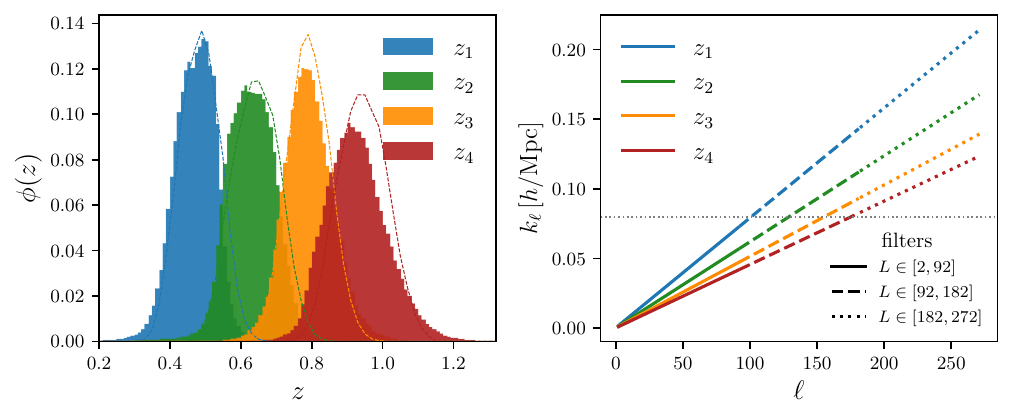}
  \caption{{\bf Redshift distribution and scale cuts.} \textit{Left}: redshift distribution for each tomographic bin used here (filled histograms). The distribution of the simulated \absum sample is shown as dashed lines. \textit{Right}: multipole-wavenumber relation, in the Limber approximation, for the filtered bispectrum data. The gray horizontal line marks the fiducial scale cut used here for the \fsb measurements.}
\label{fig:nz_and_filters}
\end{figure}

    We also use galaxy overdensity maps constructed from the photometric luminous red galaxy (LRG) sample from the 9th data release (DR9) of the Legacy Survey \cite{dr9_desi_2023}. Specifically, we use the galaxy overdensity maps and associated masks made publicly available by \citep{zhouDESILuminousRed2023}, and used in the analysis of \cite{sailer_cosmological_2024} (S25 hereafter). We do not perform any further post-processing steps on these maps, which are provided natively at our chosen resolution $N_{\rm side}=2048$. Maps are provided for four different redshift bins, spanning the redshift range $0.3\lesssim z\lesssim1.3$. The redshift distributions of each bin are also publicly available, and shown for completeness in the left panel of Fig. \ref{fig:nz_and_filters}. We use these redshifts to build our theoretical predictions for the angular power spectra and bispectra. The final ingredient needed for this analysis is an estimate of the slope in the cumulative number counts of sources as a function of magnitude $s_\mu$, which we use to quantify the impact of magnification bias. We use the values of $s_\mu$ listed in Table 1 of S25. For further details regarding the sample selection and characterisation, the derivation of contaminant weights, redshift distributions, and number count slopes, we refer readers to \cite{zhouDESILuminousRed2023} and S25.

  \subsection{Power spectrum and bispectrum estimation}\label{ssec:data.cl}
    We estimate the galaxy-galaxy angular power spectrum $C_\ell^{gg}$ in each redshift bin, as well as its cross-correlation with CMB lensing convergence, $C_\ell^{g\kappa}$, using the pseudo-$C_\ell$ algorithm \citep[PCL,][]{wandelt_PCL_2001} as implemented in {\tt NaMaster} \citep{alonsoUnifiedPseudoC_2019}. All power spectra were estimated using bandpowers of width $\Delta\ell=9$, in the range $2\leq \ell\leq 6140$. The impact of residual mode coupling due to the presence of sky masks is forward-modelled in the theoretical predictions using the bandpower window functions predicted by {\tt NaMaster}. As per the procedure laid out in Appendix C of S25, an additional correction must be performed on the galaxy-convergence cross-spectrum in order to account for different masking procedures and inhomogeneous normalization introduced in the making of the CMB lensing convergence map $\kappa$. These effects are corrected using a simulation-based transfer function, computed over 480 simulated CMB lensing reconstructions \cite{2309.05659}. Once computed, we divide the $C_\ell^{g\kappa}$ measurements by the transfer function (whose inverse is given by Eq. C.1 in S25). 

    We also estimate the angular galaxy bispectrum in each redshift bin, using the filtered-squared bispectrum (\fsb) estimator presented in \cite{harscouet_fast_2025}. In this framework, the bispectrum of a given field $\delta_g$ is calculated as the cross-correlation between $\delta_g$ and the square of a filtered version of itself $(\delta_{g,L})^2\equiv(W_L\otimes\delta_g)^2$, where the filter $W_L$ suppresses all modes outside a range of multipoles around the angular scale $L$. The \fsb is thus defined as
    \begin{equation}
      \Phi^{ggg}_{LL\ell}\equiv \langle (\delta_{g,L})^2_{\ell m}\,\delta^*_{g,\ell m}\rangle,
    \end{equation}
    and is related to the reduced bispectrum $b^{ggg}_{\ell_1\ell_2\ell_3}$ via
    \begin{equation}\label{eq:fsb2blll}
      \Phi^{ggg}_{LL\ell}=\sum_{\ell_1\ell_2\ell_3}K_{LL\ell}^{\ell_1\ell_2\ell_3}\,b^{ggg}_{\ell_1\ell_2\ell_3},
    \end{equation}
    where the \fsb kernel $K^{\ell_1\ell_2\ell_3}_{LL\ell}$ is
    \begin{equation}\label{eq:fsb_kernel}
      K^{\ell_1\ell_2\ell_3}_{LL\ell_b}\equiv h^2_{\ell_1\ell_2\ell_3}\frac{W^L_{\ell_1}W^L_{\ell_2}W^{\ell_b}_{\ell_3}}{N_{\ell_b}},
    \end{equation}
    where we have explicitly noted that we use bandpowers, labelled by $\ell_b$, with $N_{\ell_b}$ the total number of modes in the bandpower ($N_{\ell_b}\equiv \sum_{\ell\in\ell_b}(2\ell+1)\,W^{\ell_b}_\ell$). Finally, the prefactor $h^2_{\ell_1\ell_2\ell_3}$ is
    \begin{equation}
      h^2_{\ell_1\ell_2\ell_3}\equiv\frac{(2\ell_1+1)(2\ell_2+1)(2\ell_3+1)}{4\pi}
      \left(\begin{array}{lll}
             \ell_1 & \ell_2 & \ell_3 \\
             0 & 0 & 0
            \end{array}
      \right)^2.
    \end{equation}

    Note that the cross-correlation between $(\delta_{g,L})^2$ and $\delta_g$ is estimated using the pseudo-$C_\ell$ method to account for mask-related mode coupling. As shown in \cite{harscouet_fast_2025}, the residual mode coupling due to carrying out the filtering operation on an incomplete sky is negligibly small. Thus defined, the \fsb therefore recovers the angular bispectrum for triangular configurations that are close to isosceles, although, as noted in \cite{harscouet_fast_2025}, this is nevertheless able to capture a large fraction of all available triangles. The estimator can also be easily generalised to capture arbitrary triangle configurations, as well as correlations between different fields \cite{harscouet_constraints_2025}.

    The filter $W_L$ defined above can take a variety of forms; in this study, we define three equal-width top-hat filters which span the range $2 \leq \ell \leq 272$, each $\Delta L = 90$ wide. For each filter $W_{L}$, there is a corresponding \fsb measurement over the range $2 \leq \ell \leq 6140$, binned in $682$ bandpowers of width $\Delta \ell = 9$. The data vector therefore contains the $gg$ and $g\kappa$ power spectra, as well as a few \fsb measurements -- the number of \fsb included in the data vector will depend on the scale cuts we implement in the analysis. Indeed, a \fsb whose filter selects scales beyond the regime where our theoretical predictions are reliable should be discarded; the right panel of Fig. \ref{fig:nz_and_filters} shows the redshift-dependent relation between Fourier modes (where we define theoretical scale cuts) and harmonic multipoles (where the filters $W_{L}$ are defined) which we use to inform our filter selection in each redshift bin. For instance, for a scale cut at $k_\text{max} \sim 0.08 h/\text{Mpc}$ (shown as the horizontal dotted line in Fig. \ref{fig:nz_and_filters}), it is safe to include the first \fsb measurement for the lowest redshift bin, and the first two measurements for the highest redshift bin. 
    
    The covariance of the data vector is estimated using the procedure first described in \cite{harscouet_fast_2025} -- for a detailed account of the various terms entering the estimate of the covariance for cross-correlations, please refer to Appendix C of H25. It is a fully analytical approximation of the covariance, easily estimated from the data itself in a model-independent way. Reinterpreting the bispectrum as a power spectrum of fields $\delta_g$ and $\delta_{g,L}^2$ allows us to reuse already-existing covariance estimation methods for power spectra \citep[see e.g.][]{cg2_PCL_covariance_2019}; in doing so (which means, in practice, simply implementing the Gaussian power spectrum covariance from \nmt), we recover all Gaussian and non-Gaussian diagonal contributions of the \fsb and \fsb-PCL covariance. Additional non-negligible off-diagonal contributions, called $N_{222}$ and $N_{32}$ for the \fsb auto-covariance and \fsb-PCL cross-covariance respectively, can be derived analytically, and are rescaled by the inverse of the effective sky fraction (to account for mode loss) before being added to the diagonal ``Gaussian'' covariance. In this analysis, we only include the $N_{222}$ term and neglect the $N_{32}$ term. This is because the latter, which is considerably smaller in amplitude than $N_{222}$ in the $\left[ \Phi_{LL\ell}^{ggg}, C_\ell^{gg} \right]$ covariance block, and simply negligible for the $\left[ \Phi_{LL\ell}^{ggg}, C_\ell^{g\kappa} \right]$ block (see Appendix C of \cite{harscouet_constraints_2025}), takes longer to compute as it contains a generalized \fsb, $\Phi_{L\ell \ell}$, in which one of the legs entering the squared field is filtered on a range of scales corresponding to the bandpowers. As shown in \cite{harscouet_fast_2025}, adding $N_{32}$ to the $N_{222}$ correction only marginally affects the \fsb residual distribution.

\section{Methodology}\label{sec:meth}
  \subsection{Simulations}\label{ssec:meth.sims}
    In order to validate the theoretical model described below, and quantify the range of scales over which we may be able to use it safely, we make use of simulation-based mock observations mimicking the data we use here.

    \absum is a suite of cosmological $N$-body simulations designed to meet and exceed the Cosmological Simulation Requirements of the DESI survey \citep{2021MNRAS.508.4017M}. The simulations were run with \textsc{Abacus} \citep{2019MNRAS.485.3370G,2021MNRAS.508..575G}, a high-accuracy, high-performance cosmological $N$-body code optimized for GPU architectures and large-volume runs, on the Summit supercomputer at the Oak Ridge Leadership Computing Facility. In this work, we use the halo light cone catalogs from the \texttt{huge} resolution box, which spans $7.5\,\gpcoh$ on a side with $8640^3$ particles, each with mass 5.7 10$^{10}$ $h^{-1}M_\odot$ (corresponding to a particle mass 27 times larger than the \texttt{base} resolution). These catalogs include the past light cone geometry needed for mock survey generation, and allow us to cover the full DESI footprint with sufficient volume \citep{2110.11413}. The simulations adopt the fiducial \textit{Planck} 2018 cosmology ($\Omega_b h^2 = 0.02237$, $\oc=0.12$, $h=0.6736$, $10^9\,A_s=2.0830$, $n_s=0.9649$).

    To mimic the DESI Extended photometric LRG sample from Data Release 10 (DR10) \citep{2023AJ....165...58Z,zhouDESILuminousRed2023}, which is based on imaging from DECaLS on Blanco, MzLS on Mayall, and BASS on Bok, we generate galaxy mocks from the \absum halo light cones using the \texttt{AbacusHOD} package. This method populates halos with galaxies according to a halo occupation distribution (HOD), for which we adopt the best-fit parameters for the DESI 1\% LRGs at $z=0.5$ and $z=0.8$ from \cite{2306.06314}, linearly interpolated between these redshifts to provide continuous evolution across $0.3<z<1.1$. The resulting catalog is then randomly downsampled to match the observed number densities in the four redshift bins of our target dataset.
    The galaxies are binned into tomographic bins matching roughly the redshift range and distribution of the target sample. For this, each galaxy is assigned a Gaussian photometric redshift error with standard deviation $\sigma_z=0.024(1+z)$, and then binned according to the resulting photometric redshift using the same bin edges that define our target LRG sample. The resulting redshift distributions are shown in the left panel of Fig. \ref{fig:nz_and_filters}.

    We construct full-sky CMB lensing convergence maps using the density shells of the \absum \texttt{huge} light cone. The matter distribution is projected into $\sim$1000 shells of thickness $\sim 10\,h^{-1}{\rm Mpc}$ between $z=0.1$ and $z=2.3$, and mapped onto \texttt{HEALPix} pixels at $N_{\rm side}=16384$ (corresponding to $0.23'$ resolution). We compute the convergence field under the Born approximation, with sources placed at $z=1089.3$ corresponding to the period of recombination.

  \subsection{Theory model}\label{ssec:meth.theory}
    \subsubsection{Angular power spectra and bispectra}\label{sssec:meth.theory.cl}
      In this work we make use of two projected tracers of the large-scale structure: the angular overdensity of LRGs $\delta_g$ in different redshift bins, and maps of the CMB lensing convergence $\kappa$.

      The angular galaxy overdensity, accounting for the impact of lensing magnification, can be related to the three-dimensional matter and galaxy overdensities ($\Delta_g$ and $\Delta_m$, respectively) via
      \begin{equation}\label{eq:delta_g_2D}
        \delta_g(\nv)=\int d\chi\,\left[q_g(\chi)\Delta_g(\chi\nv,z(\chi))+q_\mu(\chi)\Delta_m(\chi\nv,z(\chi))\right],
      \end{equation}
      where $\chi$ is the radial comoving distance, $\nv$ is a line-of-sight unit vector, and the radial kernels associated with intrinsic clustering ($q_g$) and magnification ($q_\mu$) are
      \begin{align}
        &q_g(\chi)=H(z(\chi))\,p(z(\chi)),\\
        &q_\mu(\chi)=\frac{3}{2}H_0^2\Omega_m\,(1+z)\,\chi\,\int_{z(\chi)}^\infty dz'\,p(z')\,\left(5s(z')-2\right)\left(1-\frac{\chi}{\chi(z')}\right).
      \end{align}
      Here $p(z)$ is the redshift distribution of the galaxy sample, $\Omega_m$ and $H_0$ are the matter fraction and the current expansion rate, respectively, and $s(z)$ is the slope of the cumulative number count of sources as a function of magnitude at the magnitude limit \cite{zhouDESILuminousRed2023}. Following S25, we will assume that $s$ does not vary significantly within the redshift bin, and adopt the values presented in \cite{2111.09898}. It is worth noting that, since for this sample $s>0.4$, magnification leads to a \emph{positive} contribution to the observed galaxy overdensity (i.e. regions of positive matter overdensity will contribute towards a positive $\delta_g$).

      The CMB lensing convergence traces the matter overdensities directly via
      \begin{equation}
        \kappa(\nv)=\int d\chi\,q_\kappa(\chi)\,\Delta_m(\chi\nv,z(\chi)),\hspace{12pt}
        q_\kappa(\chi)\equiv\frac{3}{2}H_0^2\Omega_m\,(1+z)\,\chi\,\left(1-\frac{\chi}{\chi_{\rm LSS}}\right),
      \end{equation}
      where $\chi_{\rm CMB}$ is the distance to the last-scattering surface at $z_{\rm CMB}\simeq1089$. Note that this expression, as well as the description of magnification used above, neglects a scale-dependent prefactor in harmonic space, $f_\ell\equiv\ell(\ell+1)/(\ell+1/2)^2$, which is negligibly different from 1 on the scales used here.

      Within the Limber approximation adopted here, the angular power spectra of $\delta_g$ and $\kappa$ can be related to the 3D power spectra of $\Delta_g$ and $\Delta_m$ via
      \begin{align}\label{eq:cl_gg}
        &C_\ell^{gg}=\int \frac{d\chi}{\chi^2}\left[q_g^2(\chi)P_{gg}(k_\ell,z(\chi))+2q_g(\chi)q_\mu(\chi)P_{gm}(k_\ell,z(\chi))+q_\mu^2(\chi)P_{mm}(k_\ell,z(\chi))\right],\\\label{eq:cl_gk}
        &C_\ell^{g\kappa}=\int\frac{d\chi}{\chi^2}q_\kappa(\chi)\left[q_g(\chi)P_{gm}(k_\ell,z(\chi))+q_\mu(\chi)P_{mm}(k_\ell,z(\chi))\right],
      \end{align}
      where $P_{gm}(k,z)$ and $P_{gg}(k,z)$ are the 3D galaxy-matter and galaxy-galaxy power spectra, respectively, and $k_\ell\equiv(\ell+1/2)/\chi$. At this point it is worth noting that the impact of lensing magnification is negligibly small in the galaxy auto-correlation, but becomes relevant at the $\sim10\%$ level in the CMB lensing cross-correlation, leading to parameter shifts of order $\sim1\sigma$ if neglected.
      
      For simplicity, the expressions above have assumed the so-called Limber approximation \cite{Limber_1953}, which is a rather good approximation given the relatively wide redshift bins used here. Nevertheless, we will calculate all power spectrum templates described in Section \ref{ssec:meth.emu} without adopting this approximation on scales $\ell<100$, where non-Limber effects are most relevant \cite{sailer_cosmological_2024}. Another potentially important large-scale effect is redshift-space distortions (RSDs) \cite{astro-ph/9708102}. In the context of projected clustering, RSDs cause a spatially-correlated migration of galaxies in and out of the selected redshift bin, leading to additional large-scale power \cite{tanidis_developing_2019}. We incorporated the impact of linear RSDs in the galaxy auto-correlation, where it is most relevant. As we show in Appendix \ref{app:tests.rsd}, the impact of these large-scale effects is relatively small, and our results are consistent with those found after removing the scales that are most affected by them.

      Finally, our analysis will incorporate measurements of the projected galaxy bispectrum $b_{\ell_1\ell_2\ell_3}^{ggg}$, related to the \fsb estimator through Eq. \ref{eq:fsb2blll}. The projected and three-dimensional bispectra are then related via the Limber integral
      \begin{equation}
        b_{\ell_1\ell_2\ell_3}^{ggg}=\int \frac{d\chi}{\chi^4}q^3_g(\chi)\,B_{ggg}(k_{\ell_1},k_{\ell_2},k_{\ell_3},z(\chi)).
      \end{equation}
      As mentioned above, the impact of magnification is negligible in the galaxy auto-spectrum and, for the same reason, we will ignore its contribution to the galaxy bispectrum. This is further justified by the significantly larger statistical uncertainties of the bispectrum measurements.

    \subsubsection{Shot noise}\label{sssec:meth.theory.shot}
      We add a shot noise contribution to the model galaxy auto-spectrum of the form
      \begin{equation}
        C_\ell^{gg,{\rm SN}}=\frac{1+\alpha_P}{\bar{n}},
      \end{equation}
      where $\bar{n}$ is the angular number density of sources, and $\alpha_P$ is a free parameter characterising deviations from purely Poissonian shot noise. This parameter thus allows us to account for other stochastic contributions to the small-scale clustering of galaxies that are effectively uncorrelated (and hence leading to a white noise component) within the scales studied.

      We will include a similar contribution to the projected bispectrum, of the form
      \begin{equation}
        b^{ggg,{\rm SN}}_{\ell_1\ell_2\ell_3}=\left(\frac{1+\alpha_P}{\bar{n}}\right)^2+\frac{1+\alpha_P}{\bar{n}}\left[C^{gg(1)}_{\ell_1}+C^{gg(1)}_{\ell_2}+C^{gg(1)}_{\ell_3}\right],
      \end{equation}
      where $C_\ell^{gg(1)}$ is the tree-level galaxy power spectrum without shot noise. Note that our model assumes implicitly that the impact of non-Poissonian shot noise can be taken into account in both the 2- and 3-point correlators by simply scaling the effective number density of sources assuming Poisson statistics ($\bar{n}_{\rm eff}=\bar{n}/(1+\alpha_P)$). Although this is not the most general model describing stochastic bias components in galaxy clustering, we find this approximation to be sufficiently accurate in this analysis. 

    \subsubsection{Three-dimensional clustering}\label{sssec:meth.theory.bias}
      The final ingredient needed in our model is a description of the three-dimensional clustering statistics of the galaxy and matter overdensities, namely $P_{gg}(k,z)$, $P_{gm}(k,z)$, and $B_{ggg}(k_1,k_2,k_3,z)$.

      Here we will use a perturbative bias expansion in Eulerian space, within the so-called Effective Field Theory of LSS \cite{astro-ph/0609413,2212.08488}. Specifically, we describe the galaxy and matter overdensities up to third order as
      \begin{equation}
        \Delta_m=\delta^{(1)}+\delta^{(2)}+\delta^{(3)},\hspace{12pt}
        \Delta_g=
        b_1\Delta_{m}+
        \frac{b_2}{2}(\delta^{(1)})^2+\frac{b_s}{2}s^2+\frac{b_{3nl}}{2}\psi_{nl}+\frac{b_{k^2}}{2}\nabla^2\delta^{(1)},
      \end{equation}
      where $\delta^{(1)}$ is the linear matter overdensity, and $\delta^{(n)}$ is the contribution to the matter overdensity at $n$-th order in standard perturbation theory. The squared traceless tidal field $s^2$ is defined as
      \begin{equation}
        s^2\equiv s_{ij}s^{ij},\hspace{12pt}s_{ij}=(\partial_i\partial_j\nabla^{-2}-\delta^K_{ij}\delta^{(1)}),
      \end{equation}
      and the third-order operator $\psi_{nl}$ is defined in e.g. \cite{0902.0991}.

      We model the galaxy-galaxy and galaxy-matter power spectra at one-loop order in standard perturbation theory (SPT), including EFT counterterm corrections:
      \begin{subequations}\label{eq:P_1l_expansion}
        \begin{align}\nonumber
          P_{gg}(k)=&\,b_1^2 P_{\rm 1-loop}(k)+
          b_1 b_2P_{\delta^{(2)}(\delta^{(1)})^2}(k)+
          b_1 b_sP_{\delta^{(2)}s^2}(k)+
          b_1 b_{3nl}P_{\delta^{(1)}\psi_{nl}}(k)\\
          &+\frac{b_2^2}{4}P_{(\delta^{(1)})^2(\delta^{(1)})^2}(k)+\frac{b_s^2}{4}P_{s^2s^2}(k)+\frac{b_2b_s}{2}P_{s^2(\delta^{(1)})^2}(k)+b_1b_{k^2}P_{\delta^{(1)}\nabla^2\delta^{(1)}}(k)\\
          P_{gm}(k)=&\,b_1P_{\rm 1-loop}(k)+\frac{b_2}{2}P_{\delta^{(2)}(\delta^{(1)})^2}(k)+\frac{b_s}{2}P_{\delta^{(2)}s^2}(k)+\frac{b_{3nl}}{2}P_{\delta^{(1)}\psi_{nl}}(k)\nonumber\\
          &+\left(\frac{b_{k^2}}{2}+b_1c_m\right)P_{\delta^{(1)}\nabla^2\delta^{(1)}}(k)\\
          P_{mm}(k)=&\,P_{\rm 1-loop}(k)+2c_mP_{\delta^{(1)}\nabla^2\delta^{(1)}}(k).
        \end{align}
      \end{subequations}
      Here $P_{\rm 1-loop}(k)$ is the 1-loop SPT matter power spectrum, and all other power spectra of the form $P_{{\cal O}_1{\cal O}_2}(k)$ are the power spectra between operators ${\cal O}_1$ and ${\cal O}_2$ at lowest non-zero order in perturbation theory. For combinations involving the higher derivatives term, we only retain the leading order contribution given by $\langle\delta^{(1)}\nabla^2\delta^{(1)}\rangle$, ignoring the contributions from $\langle (\delta^{(1)})^2 \nabla^2\delta^{(1)} \rangle, \langle s^2 \nabla^2\delta^{(1)} \rangle$ and $\langle \nabla^2\delta^{(1)} \nabla^2\delta^{(1)} \rangle$ \citep{1611.09787}. Note that we have added a counterterm for the matter field proportional to $k^2$, with a free parameter $c_m$. A similar counterterm would also be present for the galaxy overdensity, although its amplitude is fully degenerate with the non-local bias parameter $b_{k^2}$.

      In turn, we model the galaxy bispectrum at tree level in SPT, in which case:
      \begin{equation}\label{eq:Bggg-theory}
        B_{ggg}(k_1,k_2,k_3)=2P(k_1)P(k_2)\left[b_1^3F_2({\bf k}_1,{\bf k}_2)+\frac{b_1^2b_2}{2}Q_2({\bf k}_1,{\bf k}_2)+\frac{b_1^2b_s}{2}T_2({\bf k}_1,{\bf k}_2)\right]+({\rm c.c.}),
      \end{equation}
      where ``c.c.'' stands for the other two cyclic permutations of the three wavevectors, $P(k)\equiv P_{\delta^{(1)}\delta^{(1)}}(k)$ is the linear matter power spectrum, $F_2({\bf k}_1,{\bf k}_2)$ is the second-order density SPT kernel, and $Q_2$ and $T_2$ are the corresponding kernels for the quadratic and tidal fields:
      \begin{equation}
        F_2({\bf k}_1,{\bf k}_2)\equiv\frac{5}{7}+\frac{1}{2}\left(\frac{k_1}{k_2}+\frac{k_2}{k_1}\right)+\frac{2}{7}\mu_{12}^2,
        \hspace{6pt}Q_2({\bf k}_1,{\bf k}_2)\equiv 1,
        \hspace{6pt}T_2({\bf k}_1,{\bf k}_2)\equiv\mu_{12}^2-\frac{1}{3},
      \end{equation}
      with $\mu_{12}\equiv {\bf k}_1\cdot{\bf k}_2/k_1k_2$. Note that we do not include non-local or higher-derivative contributions to the bispectrum, since these are mostly relevant in the presence of non-linear redshift-space distortions and at one-loop order \cite{2302.04414}.

  \subsection{Accelerating theory predictions}\label{ssec:meth.emu}
    Although the calculation of the theoretical predictions for the angular power spectra can be carried out with relative efficiency, particularly using fast schemes to calculate PT convolutions, such as {\tt FAST-PT} \cite{1603.04826,1609.05978}, the time to complete a fully converged Markov chain (see Section \ref{ssec:meth.like}) can still be significant (e.g. hours or days). This problem is exacerbated in the case of the angular bispectrum, given the large number of Limber integrals needed to cover all required triangle configurations, and the additional complication of convolving the result with the \fsb kernel (Eq. \ref{eq:fsb_kernel}). In order to increase the efficiency of our theory predictions as a function of cosmological parameters, we employ machine-learning based emulators. Concretely, we will emulate projected quantities, integrated over the radial kernels of the tracers under study. Thus, our emulator is strictly tied to the specific dataset considered in this work.
   
    In both the 2-point and 3-point observables studied here, the bias coefficients and counterterms ($b_1$, $b_2$, $b_s$, $b_{3nl}$, $b_{k^2}$, $c_m$) are always multiplicative and can be brought outside the Limber integrals. For instance, consider the galaxy-convergence power spectrum in Eq. \ref{eq:cl_gk}. Within the perturbative bias expansion used here, it can be written as
    \begin{align}
      C_\ell^{g\kappa}=&b_1{\sf C}^{g\kappa}_{{\rm 1-loop},\ell}+\frac{b_2}{2}{\sf C}^{g\kappa}_{\delta^{(2)}(\delta^{(1)})^2,\ell}+\frac{b_s}{2}{\sf C}^{g\kappa}_{\delta^{(2)}s^2,\ell}+\frac{b_{3nl}}{2}{\sf C}^{g\kappa}_{\delta^{(1)}\psi_{nl},\ell}\\
      &+\left(\frac{b_{k^2}}{2}+b_1c_m\right){\sf C}^{g\kappa}_{\delta^{(1)}\nabla^2\delta^{(1)},\ell}+{\sf C}^{\mu\kappa}_{{\rm 1-loop},\ell}+c_m{\sf C}^{\mu\kappa}_{\delta^{(1)}\nabla^2\delta^{(1)},\ell},
    \end{align}
    where we have defined the angular power spectrum templates corresponding to the different PT power spectra in Eq. \ref{eq:P_1l_expansion}:
    \begin{equation}
      {\sf C}^{ab}_{{\cal O}_1{\cal O}_2,\ell}\equiv\int\frac{d\chi}{\chi^2}q_a(\chi)q_b(\chi)\,P_{{\cal O}_1{\cal O}_2}(k_\ell,z(\chi)).
    \end{equation}
    A similar expression can be derived for $C_\ell^{gg}$, with the addition of the shot noise component. Crucially, these templates depend only on a small number of cosmological parameters (two in our case, see below), and can therefore be emulated a priori.

    In the same way for the galaxy bispectrum, where the acceleration enabled by emulators is even more essential, we may write
        \begin{equation}
\Phi^{ggg}_{LL\ell}=b_1^3{\sf F}^{ggg}_{LL\ell}+\frac{b_1^2b_2}{2}{\sf Q}^{ggg}_{LL\ell}+\frac{b_1^2b_s}{2}{\sf T}^{ggg}_{LL\ell}+b_1^2(1+\alpha_P){\sf SN}^{ggg}_{2,LL\ell}+(1+\alpha_P)^2{\sf SN}^{ggg}_{1,LL\ell} \, ,
    \end{equation}
    where we have defined the \fsb templates:
    \begin{align}\label{eq:FSB_ell_sum}
      {\sf X}^{ggg}_{LL\ell}\equiv\sum_{\ell_1\ell_2\ell_3}K^{\ell_1\ell_2\ell_3}_{LL\ell}{\sf b}^{ggg,X}_{\ell_1\ell_2\ell_3},
    \end{align}
    with the \fsb kernel defined in Eq. \ref{eq:fsb_kernel}, and
    \begin{align}\label{eq:fsb_templates}
      &{\sf b}^{ggg,X}_{\ell_1\ell_2\ell_3}\equiv\int\frac{d\chi}{\chi^4}q^3_g(\chi)\left[2P(k_{\ell_1})P(k_{\ell_2})X({\bf k}_{\ell_1},{\bf k}_{\ell_2})+({\rm c.c.})\right]\hspace{12pt} {\rm for}\,\,X\in\{F,Q,T\},\\
      &{\sf b}^{ggg,{\rm SN}_2}_{\ell_1\ell_2\ell_3}\equiv\frac{1}{\bar{n}}\left[{\sf C}^{gg}_{\delta^{(1)}\delta^{(1)},\ell_1}+{\sf C}^{gg}_{\delta^{(1)}\delta^{(1)},\ell_2}+{\sf C}^{gg}_{\delta^{(1)}\delta^{(1)},\ell_3}\right],\hspace{12pt} {\sf b}^{ggg,{\rm SN}_1}\equiv\frac{1}{\bar{n}^2}
    \end{align}
    As before, the ${\sf X}$ templates depend only on cosmological parameters. 

    We build one emulator per template and per redshift bin. The architecture of each of our emulators is a Feed-Forward Neural Network, with 2 hidden layers of 128 nodes each. The input layer of the network has size 2, corresponding to $\{\log (10^{10} A_{\rm s}), \Omega_{\rm c}h^2\}$. The output layer has the same size as our $\ell$ vector. For each emulator, we create a sample of 10,000 input-output pairs, which we split into a training set (holding 90\% of the sample) and a test set (with the remaining 10\%). We build and train our models with \texttt{Keras3} \citep{chollet2015keras} and \texttt{TensorFlow} \citep{tensorflow2015-whitepaper}, using a Rectified Linear Unit (\texttt{ReLU}) activation function, and optimizing our weights and biases with the \texttt{Adam} gradient descent algorithm \citep{kingma2014adam}, with exponentially decreasing learning rate. We use mean squared errors between the training and test sets as a loss function and, during training, also compute an accuracy metric that accounts for the percentage of output nodes differing less than 0.5\% from the ground truth. We use this metric to establish an early stopping criterion. The emulators we obtain with this procedure are able to reproduce the corresponding test sets at better than 1\% accuracy on all of the considered scales, see App. \ref{ssec:em.acc}.

    The numerical computation of the templates appearing in Eq. \eqref{eq:fsb_templates}, with the method developed in H25, takes $\approx5\,\mathrm{s}$ when already emulating the linear power spectrum \cite{2104.14568}, rather than calling the Einstein-Boltzmann solver, and interpolating the evaluation of Eq. \eqref{eq:FSB_ell_sum}. Still, this makes posterior sampling, which we find requires evaluating the likelihood $\mathcal{O}\left(10,000 - 100,000 \right)$ times, practically unfeasible, especially when combining the redshift bins. Our approach of emulating the templates reduces the single evaluation time of Eq.\eqref{eq:fsb_templates} by three orders of magnitude, down to $\approx 5\, \mathrm{ms}$, enabling a feasible, and extremely fast, exploration of the posterior with MCMC methods.

  \subsection{Likelihood}\label{ssec:meth.like}
    \renewcommand{\arraystretch}{1.5}
    \begin{table}
    \centering
    \begin{tabular}{|l|l|}
    \hline
    \textbf{Parameter} & \textbf{Prior} \\
    \hline
    $\Omega_c h^2$ & $\mathcal{U}(0.1, 0.14)$ \\
    $\log\left(10^{10}A_s\right)$ & $\mathcal{U}(2.3, 3.5)$  \\
    \hline
    $b_1$ & $\mathcal{U}(1.2,3.2)$  \\
    $b_2$ & $\mathcal{U}(-5,5)$  \\
    $b_s$ & $\mathcal{U}(-10,10)$   \\
    \hline
    \end{tabular}
    \hspace{0.5cm}
    \begin{tabular}{|l|l|}
    \hline
    \textbf{Parameter} & \textbf{Prior} \\
    \hline
    $b_{3nl}$ & $\mathcal{U}(-10,10)$   \\
    $b_{k^2}$ & $\mathcal{N}(0,15\,\mathrm{Mpc}^{2})$   \\
    $c_{m}$ & $\mathcal{N}(0,10\,\mathrm{Mpc}^{2})$  \\
    \hline
    $s_\mu$ & $\mathcal{N}(s_\mu^\mathrm{fid},0.1)$  \\
    $\alpha_{P}$ & $\mathcal{N}(0,0.3)$\\
    \hline
    \end{tabular}
    \caption{Model parameters and priors used.}
    \label{tab:parameter_table}
    \end{table}
    We derive parameter constraints from the measurements described in Section \ref{sec:data} assuming a Gaussian likelihood of the form
    \begin{equation}
      -2\log p({\bf d}|\vec{\theta})=({\bf d}-{\bf t}(\vec{\theta}))^T{\sf C}^{-1}({\bf d}-{\bf t}(\vec{\theta}))+K,
    \end{equation}
    where ${\bf d}$ is the data vector, ${\bf t}(\vec{\theta})$ is the theory prediction, described in the preceding sections and dependent upon a set of free parameters $\vec{\theta}$, and $K$ is an arbitrary normalisation constant.

    In our analysis, ${\bf d}$ will comprise a set of power spectrum ($C^{gg}_\ell$ and/or $C^{g\kappa}_\ell$) and \fsb measurements ($\Phi^{ggg}_{LL\ell}$), with the covariance matrix estimated as described in Section \ref{ssec:data.cl}. The range of scales over which different power spectra and bispectra are used was determined following the validation against simulations presented in Section \ref{ssec:res.sims}. In our fiducial analysis we will use power spectrum data on angular scales $\ell<\ell_{\rm max}^P\equiv k_{\rm max}^P\,\chi(\bar{z})$, where $\chi(\bar{z})$ is the comoving distance to the mean redshift of the sample, and $k_{\rm max}^P=0.2\,\ihMpc$. Likewise, we will include \fsb measurements on scales $(L_{\rm mean},\ell)<\ell_{\rm max}^B\equiv k_{\rm max}^B\chi(\bar{z})$, with $k_{\rm max}^B=0.08\ihMpc$. Here $L_{\rm mean}$ is the mean multipole of the top-hat filter defined in Sect. \ref{ssec:data.cl}; our fiducial criterion thus corresponds to including the first filter for all bins, as well as the second filter for $z_3$ and $z_4$. Note that we find that the constraints on $\Omega_m$ and $\sigma_8$ do not improve significantly by including smaller scales. In addition to this, we remove the largest angular scales in the power spectrum measurements, with $\ell<29$ (i.e. we discard the first three bandpowers).
    
    The free parameters of our model will be the full set of bias expansion parameters, including the counter-term and shot-noise amplitude, $\{b_1,b_2,b_s,b_{k^2},c_m,\alpha_P\}$, as well as two cosmological parameters, the physical abundance of cold dark matter $\oc$, and the amplitude of primordial scalar fluctuations $\lAs$. We fix all other $\Lambda$CDM cosmological parameters as follows: following \cite{sailer_cosmological_2024}, we use $\Omega_mh^3=0.09633$ \cite{collaboration_planck_2020} as a proxy for the angular acoustic scale, measured by \planck at very high precision, as well as $\Omega_bh^2=0.02236$, $n_s=0.9649$ (also measured by \planck). We also fix the sum of neutrino masses to $M_\nu=0.06\,{\rm eV}$. Although the parameters we will sample are $\oc$ and $\lAs$, we will most commonly present our constraints in terms of the matter fraction $\Omega_m$ and the linear matter power spectrum amplitude $\sigma_8$, as derived parameters. When considering multiple redshift bins in the same dataset, we will assume that the clustering of galaxies and matter in each bin is described by a different set of bias parameters (including $\alpha_P$ and $c_m$).

    We sample the posterior distribution using \texttt{nautilus}, a nested sampling algorithm that optimises the way its live points explore the parameter space thanks to the use of Neural Networks \cite{nautilus}. We use 1000 live points, with 4 Neural Networks, and with convergence criterion $f_\mathrm{live}=0.01$ and $n_\mathrm{eff}=5000$. To construct this posterior we impose the priors on our model parameters listed in Table \ref{tab:parameter_table}. The flat priors on $\oc$ and $\lAs$ coincide with the parameter ranges over which the emulator described in Section \ref{ssec:meth.emu} is trained, and were chosen to ensure that they enclose the region of parameter space that can be constrained by the data in each redshift bin. The priors for $b_2$, $b_s$, and $b_{3nl}$ were chosen to be sufficiently broad to encompass the ranges over which they may be constrained by two-point function data alone. The prior on $b_{k^2}$ mimics the prior used in \cite{sailer_cosmological_2024} for the counterterm amplitude and the prior on $c_m$ is based on the arguments described in \cite{ivanov_precision_2022}, whereby
    \begin{equation}
      c_m \simeq \mathcal{O}(1) \times \frac{1}{k_{NL}^2}\approx \mathcal{O}(1) \times 5 \,h^{-2}\mathrm{Mpc}^2
    \end{equation}
    for $k_{NL}(z=0.5) \simeq0.45 \,\ihMpc$.

\section{Results}\label{sec:res}
  We now present the results of our analysis, beginning with a validation of the model using simulations before applying it to the real LRG data.

  \subsection{Validation on simulations}\label{ssec:res.sims}
    Simulation data provides a controlled setting, where the true underlying cosmology is known, to assess the performance of the theoretical model employed.

    A key aspect of full-shape fits of galaxy clustering data is understanding the regime of validity of the model, i.e. defining respectively the largest and smallest scales that can be employed for the purpose of parameter inference. While the limiting factor on the largest scales is usually the presence of systematics, or the potential mismodelling of large-angle effects (e.g. the Limber approximation, redshift-space distortions, etc.), the modelling on smaller scales is limited by non-linearities in the clustering. The latter is particularly subtle to assess, and several approaches have been proposed in the literature, most of which rely on \textit{a posteriori} validation by varying the small-scale cut $k_\mathrm{max}$ in the fit. In general, we expect the fully theoretical, first-principle, model based on 1-loop (tree level) EPT to be able to describe the mildly non-linear scales in the power spectrum up to a $k_\mathrm{max}^P$ of order $\approx0.3\,\ihMpc$ ($\approx0.1\,\ihMpc$), as extensively validated for 3D clustering on N-body simulations \citep{1805.12394,2111.00501,2307.03226} for the redshifts of interest to this work.
    \begin{figure}
      \centering
      \includegraphics[width=\linewidth]{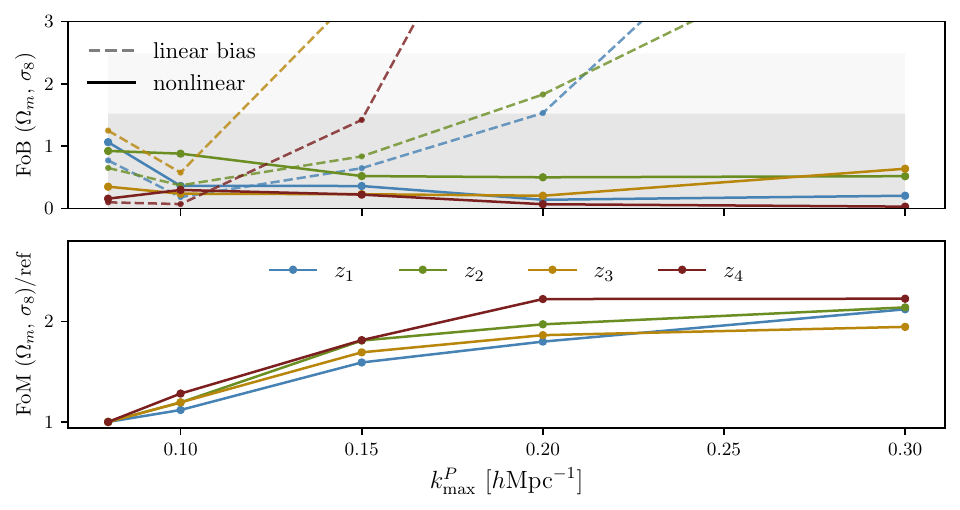}
      \caption{{\bf Model performance in simulated data.} Figure of Bias (FoB, top panel), and Figure of Merit (FoM, bottom panel) for the galaxy-galaxy and galaxy-$\kappa$ angular power spectra in each redshift bin, for the two cosmological parameters $\{\Omega_m,\sigma_8\}$, as a function of the small-scale cut $k_{\rm max}^P$. Note that the FoM is reported relative to its value for $k_{\rm max}^P=0.08\,\ihMpc$, in order to show results from all bins on the same scale. The solid lines refer to the nonlinear bias model assumed in our fiducial analysis, while dashed lines refer to a simplistic linear bias model. While linear bias can only give unbiased cosmological results up to $k_{\rm max}^P=0.1~h~\mathrm{Mpc}^{-1}$, the nonlinear model can extend to $k_{\rm max}^P=0.3~h~\mathrm{Mpc}^{-1}$ (but without significant gains in terms of extracted information for $k_{\rm max}^P>0.2~h~\mathrm{Mpc}^{-1}$). }\label{fig:FoB_FoM_sims}
    \end{figure}

    With this aim, we thus start by analysing the mock LRG catalogue and CMB lensing map generated from the \absum simulation as described in Section \ref{ssec:meth.sims}. We analyse the simulated data using the same pipeline as for the real data, measuring the galaxy auto-spectrum, the galaxy-$\kappa$ cross-spectrum, and the $ggg$ \fsb. We will then extract parameter constraints from these measurements using the covariance matrix estimated for the real data, allowing us to quantify the amplitude of any potential parameter biases against their expected measurement uncertainties in the real data. Note, however, that the simulated data vector has intrinsically smaller statistical uncertainties than the data, since the catalogue and $\kappa$ map cover the full celestial sphere, and the $\kappa$ map does not include any noise contribution from lensing reconstruction. This means that we should expect the scatter in the best-fit parameters extracted from the simulation to be much smaller than the posterior uncertainties, so any significant deviation from the true cosmology of the simulation will be likely due to the limitations of the theory model used.

    To quantify the adequacy of the theoretical model, we consider two key performance metrics commonly employed in the literature \cite{collaboration_euclid_2024}. Specifically, we compute the Figure of Bias (FoB), which quantifies the unbiasedness of the model in the recovery of the fiducial parameters, and the Figure of Merit (FoM), which measures statistical power in constraining the parameters. These are computed for the two cosmological parameters $\vec{\theta}=\{\Omega_m,\,\sigma_8\}$ as a function of the small-scale cut. The FoB and FoM are defined as
    \begin{equation}
      {\rm FoB}\equiv\sqrt{(\langle\vec{\theta}\rangle-\vec{\theta}_{\rm true})^T{\rm Cov}_\theta^{-1}(\langle\vec{\theta}\rangle-\vec{\theta}_{\rm true})},
      \hspace{12pt}
      {\rm FoM}\equiv\frac{1}{\sqrt{\det({\rm Cov}_\theta)}},
    \end{equation}
    where $\langle \vec{\theta}\rangle$ is the mean of the posterior distribution for the model parameters $\vec{\theta}$, $\vec{\theta}_{\rm true}$ is the true value of the parameter $\theta$, and ${\rm Cov}_\theta$ is the covariance of the posterior. The FoB thus quantifies the bias induced on the input parameters as a fraction of their uncertainties, while FoM quantifies the constraining power of the model.
    \begin{figure}
      \centering
      \begin{subfigure}{0.48\textwidth}
        \centering
        \includegraphics[height=8.cm]{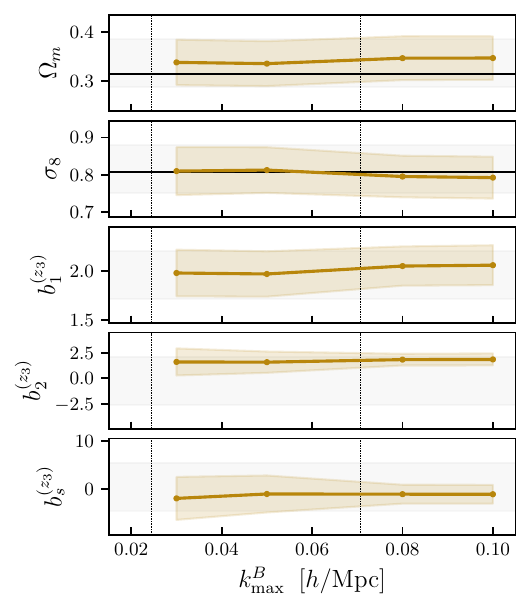}
      \end{subfigure}
      \hfill
      \begin{subfigure}{0.48\textwidth}
        \centering
        \includegraphics[height=8.cm]{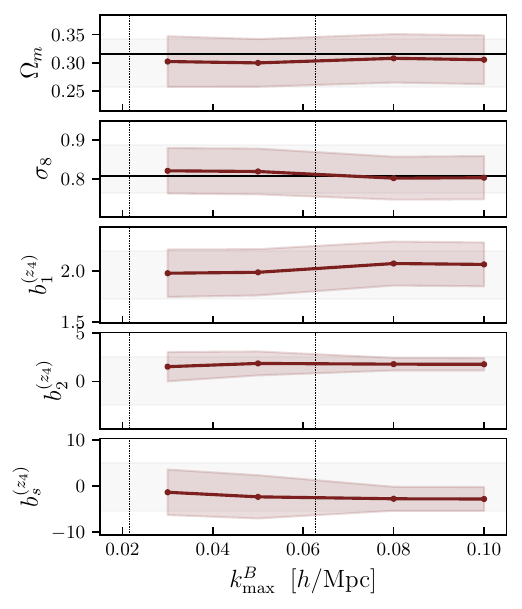}
      \end{subfigure}
      \caption{{\bf Consistency with bispectrum scale cut on simulation data.} Posterior constraints on the model parameters as a function of the small-scale cut used for the galaxy bispectrum $k_\mathrm{max}^{B}$. Results are shown for two redshift bins, $z_3$ and $z_4$ (left and right panels, respectively). Similar results are found for the first two redshift bins. The shaded coloured bands show the $68\%$ constraints on each parameter, with the gray band showing the constraints found when using only power spectrum data with $k_{\rm max}^P=0.08\,\ihMpc$ (and $\ell_{\rm min}=29$). Vertical lines represent, incrementally, the high-$\ell$ bound of the different filters used.}\label{fig:kmaxB-running}
    \end{figure}

    We begin by quantifying the scales over which the one-loop bias model used here is able to describe the projected power spectra $C_\ell^{gg}$ and $C_\ell^{g\kappa}$. For this, we analyse the simulated measurements excluding the \fsb for now. Fig. \ref{fig:FoB_FoM_sims} reports the result of the fits for each individual redshift bin. For reference, we also report the FoB for a  ``linear bias'' model in which $b_1$ is the only nuisance parameter. This simpler model, while recovering unbiased results at very large scales ($k_\mathrm{max}^{P}\lesssim0.1\,\ihMpc$), fails shortly afterwards. The apparently counter-intuitive fact that the metric, at a given scale cut, is worse for higher-redshift bins is an artifact of plotting as a function of comoving scales (as opposed to angular scales): a given $k_\mathrm{max}^{P}$ corresponds to a higher multipole at higher redshifts, and the measurements thus achieve a stronger constraining power (i.e. smaller uncertainties). In contrast, the higher-order bias model used here (see Section \ref{sec:meth}) consistently returns unbiased constraints on the two cosmological parameters for the various scale cuts and redshift bins that we consider. Furthermore, while there is a notable increase in constraining power when including mildly non-linear scales, this seemingly saturates already at $k_{\rm max}^P\sim 0.3\,\ihMpc$, or earlier ($k_{\rm max}^{P}\sim 0.2\,\ihMpc$) for the higher-redshift bins. This saturation can be related to the impact of shot noise and CMB lensing reconstruction noise on the small-scale statistical uncertainties, as well as the model flexibility allowed by the multiple bias parameters in the model, particularly the counter-term contributions proportional to $k^2P(k)$. This is explored further in Appendix \ref{app:tests.simplebias}. Given these results, we make the conservative choice of setting the same small-scale cut of $k_\mathrm{max}^{P}=0.2\,\ihMpc$ for all the redshift bins, which we apply to both $C_\ell^{gg}$ and $C_\ell^{g\kappa}$.  

    \begin{figure}[t]
      \centering
      \begin{subfigure}[t]{0.48\textwidth}
        \centering
        \includegraphics[width=\linewidth]{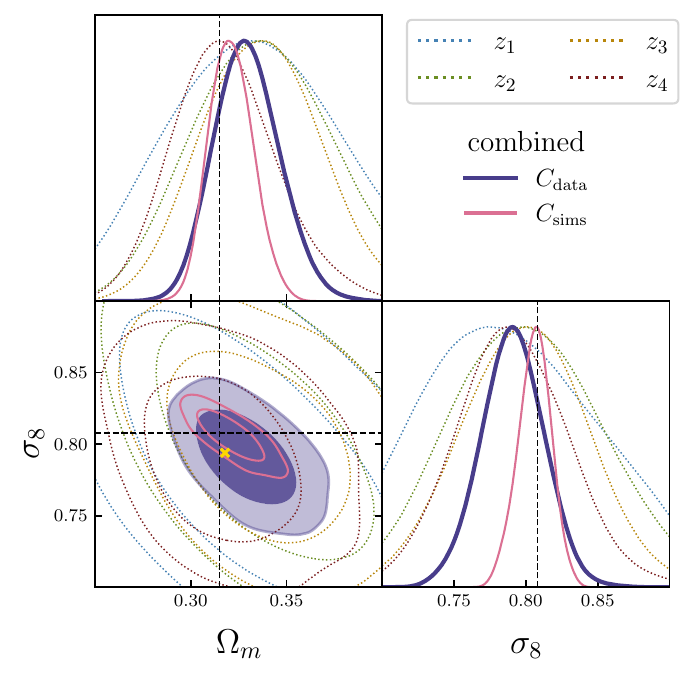}
      \end{subfigure}
      \hfill
      \begin{subfigure}[t]{0.48\textwidth}
        \centering
        \includegraphics[width=\linewidth]{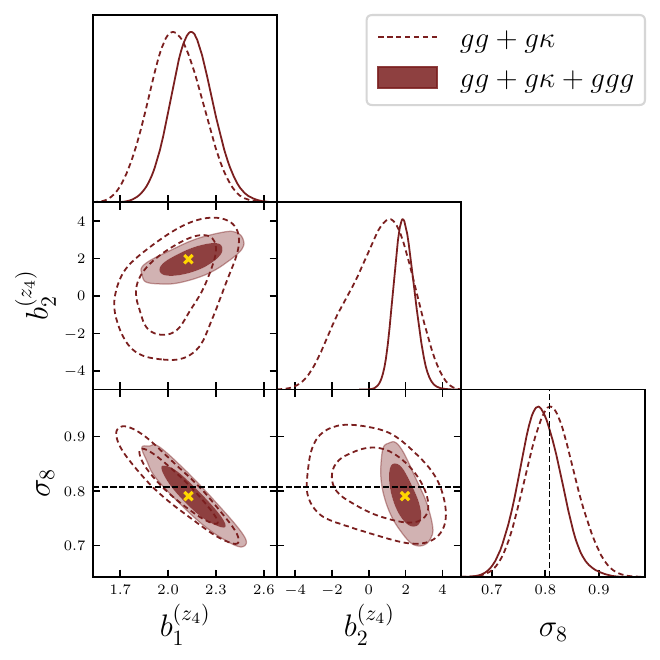}
      \end{subfigure}
      \caption{{\bf Constraints from the \absum simulated data.} \textit{Left}: posterior, for single and combined redshift bins, of the cosmological parameters in the combined $gg+g\kappa+ggg$ analysis. The dashed lines show the true cosmological parameters of the \absum simulation, while the yellow cross marks the best-fit parameters found using the data covariance in the likelihood. The solid purple contours how the constraints obtained from the combination of all redshift bins, with the dotted contours showing the constraints found from individual redshift bins, all using the data covariance. The constraints found using the simulation covariance (significantly smaller due to the larger sky area and lack of lensing reconstruction noise) are shown in pink. The true cosmological parameters are recovered in all cases with no significant bias. \textit{Right}: posterior, for the highest redshift bin, showing the relevant degeneracies of $\sigma_8$ with the bias parameters with and without the 3-point information, using the data covariance. This shows how the 1D posterior for $\sigma_8$ is affected by the marginalization over the bias parameters, especially along the degeneracy with $b_2$, when new information is added. }\label{fig:sim-fits-combined}
    \end{figure}

    To estimate a reliable scale cut for the three-point correlation function, we carry out a separate, dedicated analysis. The main difficulty in doing so is the fact that, on its own, $\Phi^{ggg}_{LL\ell}$ is not able to place meaningful constraints on cosmological and bias parameters simultaneously. For the same reason, the same metrics used above (FoB and FoM), if applied to the combination of $C_\ell^{gg}$, $C_\ell^{g\kappa}$ and $\Phi^{ggg}_{LL\ell}$, would be dominated by the two-point functions, making them almost insensitive to the bispectrum. Thus, we adopt a different strategy, combining $\Phi^{ggg}_{LL\ell}$ with $C_\ell^{gg}$ and $C_\ell^{g\kappa}$, but including the latter two only on the largest scales ($k_{\rm max}^P=0.08\,\ihMpc$), thus minimising the information extracted by the two-point function. Furthermore, rather than simply examining the FoB and FoM directly, we will instead quantify the stability of the posterior constraints on all parameters as a function of $k_{\rm max}^B$. Fig. \ref{fig:kmaxB-running} shows the constraints on all cosmological and bias parameters for the two highest redshift bins as a function of $k^B_{\rm max}$ in this setup. Similar results were found for the other two redshift bins. We find that the preferred values of both $\Omega_m$ and $\sigma_8$ are largely stable with respect to the choice of $k_{\rm max}^B$, and their statistical uncertainties only vary slowly within the range of scales studied. In turn, the constraints on the bias parameters benefit strongly from the inclusion of the three-point function. Their preferred value is also stable with respect to $k_{\rm max}^B$. Given these results, we choose a scale cut of $k_{\rm max}^B=0.08\,\ihMpc$, compatible with the values assumed in three-dimensional clustering studies \cite{1908.01774, ivanov_precision_2022}, as a compromise between the final size of the data vector and the desire to improve the constraints on the higher-order bias parameters.

    Having chosen our fiducial scale cuts, $(k_{\rm max}^P,\,k_{\rm max}^B)=(0.2,\,0.08)\,\ihMpc$, we now derive constraints from the full data vector, combining all two- and three-point correlations from the four redshift bins. The resulting constraints on $\Omega_m$ and $\sigma_8$ are shown in the left panel of Fig. \ref{fig:sim-fits-combined}. The constraints found using the data covariance matrix are compatible with the true cosmology of the \absum simulations within $1\sigma$, and the best-fit parameter values (defined as the point in the MCMC chain with the lowest $\chi^2$) are very close to the truth. Similar results are found for each redshift bin when analysed individually, and these are also shown in the figure for comparison. To determine whether the small shift observed between the posterior constraints and the true parameter values is a consequence of volume/projection effects or a genuine parameter bias, we repeat our analysis using the covariance matrix of the simulation (which, as discussed above, leads to significantly smaller data errors). The parameter constraints found in this case, shown in Fig. \ref{fig:sim-fits-combined}, are significantly tighter and perfectly compatible with the true parameter values. This suggests that the small parameter shift observed with the data covariance is indeed a result of projection effects. To confirm this further, the right panel of Fig. \ref{fig:sim-fits-combined} shows the constraints on $\sigma_8$ and the $b_1$ and $b_2$ bias parameters of the fourth redshift bins. We can see that the multi-parameter posterior distribution is clearly non-Gaussian, and marginalisation over the bias parameters, particularly in the case of $b_2$, can lead to a shift in the marginalised posterior distribution.

    In conclusion, we find that the bias model used here can describe our measurements with sufficient accuracy within our fiducial scale cuts. In Appendix \ref{app:tests.simplebias} we explored the ability of simpler bias parametrisations to recover unbiased cosmological constraints, using the mock data and validation procedure described here. The results of these simplified models will be briefly discussed in the next section.
  
  \subsection{Cosmological constraints from 2- and 3-point correlators}\label{ssec:res.cosmo}
    \begin{table}
      \renewcommand{\arraystretch}{1.1}
      \centering
      \begin{tabular}{|c|l|c|c|c|}
        \hline
        $z_\mathrm{bin}$ & \textbf{Dataset} &  $\Omega_m$ & $\sigma_8$ & fit $\chi^2$ (PTE) \\
        \hline
        \multirow{2}{*}{$z_1$} 
        & $gg+g\kappa$ & $0.296\pm0.034$ & $0.686\pm0.058$ & $1.07$ ($0.35$) \\
        & $gg+g\kappa+ggg$ & $0.294\pm0.033$ & $0.686\pm0.055$ & $0.98$ ($0.50$) \\
        \hline
        \multirow{2}{*}{$z_2$} 
        & $gg+g\kappa$ & $0.329\pm0.035$ & $0.772\pm0.050$ & $0.96$ ($0.55$) \\
        & $gg+g\kappa+ggg$ & $0.337\pm0.036$ & $0.761\pm0.046$ & $0.91$ ($0.69$) \\
        \hline
        \multirow{2}{*}{$z_3$} 
        & $gg+g\kappa$ & $0.297\pm0.025$ & $0.773\pm0.042$ & $1.02$ ($0.44$) \\
        & $gg+g\kappa+ggg$ & $0.302\pm0.024$ & $0.758\pm0.036$ & $0.99$ ($0.52$) \\
        \hline
        \multirow{2}{*}{$z_4$} 
        & $gg+g\kappa$ & $0.290\pm0.023$ & $0.785\pm0.040$ & $0.93$ ($0.65$) \\
        & $gg+g\kappa+ggg$ & $0.296\pm0.022$ & $0.765\pm0.035$ & $0.89$ ($0.78$) \\
        \hline
        \multirow{2}{*}{$z_2+z_3+z_4$}
        & $gg+g\kappa$ & $0.301\pm0.016$ & $0.774\pm0.023$ & $0.96$ ($0.64$) \\ 
        & $gg+g\kappa+ggg$ & $0.307\pm0.015$ & $0.761\pm0.020$ & $0.96$ ($0.65$) \\
        \hline
        \multirow{2}{*}{All bins}
        & $gg+g\kappa$ & $0.299\pm0.014$ & $0.759\pm0.020$ & $0.99$ ($0.54$)\\ 
        & $gg+g\kappa+ggg$ & $0.303\pm0.014$ & $0.753\pm0.018$ & $0.97$ ($0.63$) \\
        \hline
        \multirow{2}{*}{\shortstack{All bins\\ \scriptsize $\{b_1,\,b_2,\, b_s\}$}}
        & $gg+g\kappa$ & $0.299\pm0.011$ & $0.778\pm0.020$ & $0.95$ ($0.72$)\\ 
        & $gg+g\kappa+ggg$ & $0.300\pm0.010$ & $0.764\pm0.018$ & $0.95$ ($0.75$) \\
        \hline
        \multirow{2}{*}{\shortstack{All bins\\ \scriptsize $\{b_1,\,b_{k^2},\,c_m\}$}}
        & $gg+g\kappa$ & $0.295\pm0.011$ & $0.772\pm0.021$ & $0.94$ ($0.74$)\\ 
        & $gg+g\kappa+ggg$ & $0.293\pm0.009$ & $0.785\pm0.016$ & $0.98$ ($0.61$) \\
        \hline
      \end{tabular}
      \caption{\textbf{Cosmological parameter constraints} found from different data combinations. Results are shown for each redshift bin independently, as well as for the combination of the last three bins and all four bins. The last two sets of rows show the constraints found when analysing the data under the simplified bias models explored in Appendix \ref{app:tests.simplebias}. In each case, we show constraints from both two-point and two- and three-point correlators. The last column shows the reduced $\chi^2$ and probability-to-exceed for the best-fit model.}\label{tab:results_table}
    \end{table}
    
    \begin{figure}
      \centering
      \includegraphics[width=\linewidth]{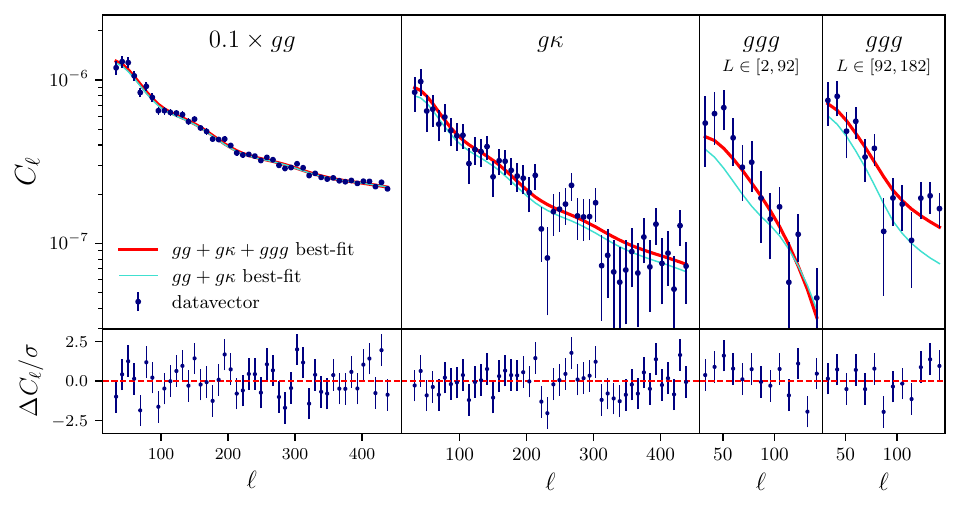}
      \caption{{\bf Power spectrum and \fsb measurements, and model fit.} Measurements of $C_\ell^{gg}$, $C_\ell^{g\kappa}$, and $\Phi_{LL\ell}^{ggg}$ in the fourth redshift bin. Results are shown over the scales used in the analysis, with the \fsb measurements covering the first two filters (the corresponding scale ranges are shown in the legend). The red line shows the best-fit model found from the combination of all data shown here, while the blue line shows the prediction found by fitting only the two-point functions. The lower panel reports the residuals with respect to the best-fit model normalized by the statistical uncertainties.}\label{fig:fit-and-chi2}
    \end{figure}
    Having validated our bias model and scale cuts, we now apply them to the analysis of the real data. Our constraints on $\Omega_m$ and $\sigma_8$ for different data combinations, as well as the best-fit $\chi^2$ and associated probability-to-exceed (PTE, i.e. the probability of obtaining a $\chi^2$ value above that observed in the data, given the number degrees of freedom of the model), are summarised in Table \ref{tab:results_table}. We discuss these results in detail in the remainder of this section.

    Figure \ref{fig:fit-and-chi2} shows our power spectrum and \fsb measurements for the fourth redshift bin, together with the best-fit theory prediction. The galaxy bispectrum is measured at high significance in the two filter scales used in the analysis (covering $2\leq\ell<92$ and $92\leq\ell<182$).  Similar results are found for the other three redshift bins. As can be seen in the figure, and as quantified in Table \ref{tab:results_table}, we find that the model is able to fit all components of the data vector well in all cases explored here, with PTE values ranging from 0.35 to 0.78.

    \begin{figure}
      \centering
      \includegraphics[width=0.8\linewidth]{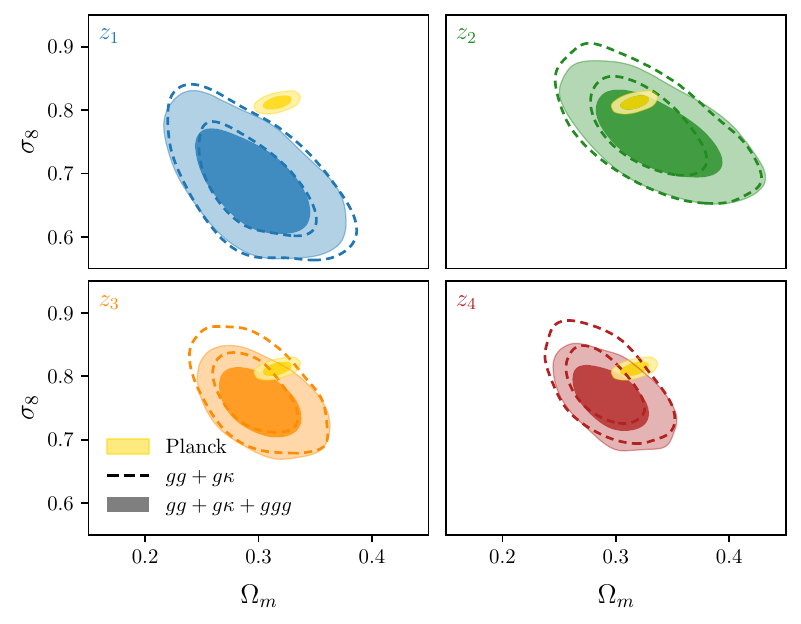}
      \caption{{\bf Cosmological constraints from each redshift bin.} Two-dimensional posterior constraint on the cosmological parameters when analysing $C_\ell^{gg}$ and $C_\ell^{g\kappa}$ (\textit{dashed lines}), and with the addition of the \fsb measurements (\textit{filled contours}). For reference, the \planck constraints are shown in yellow in all panels.}\label{fig:res_singlebins_2D}
    \end{figure}
    \begin{figure}
      \centering
        \includegraphics[width=\linewidth]{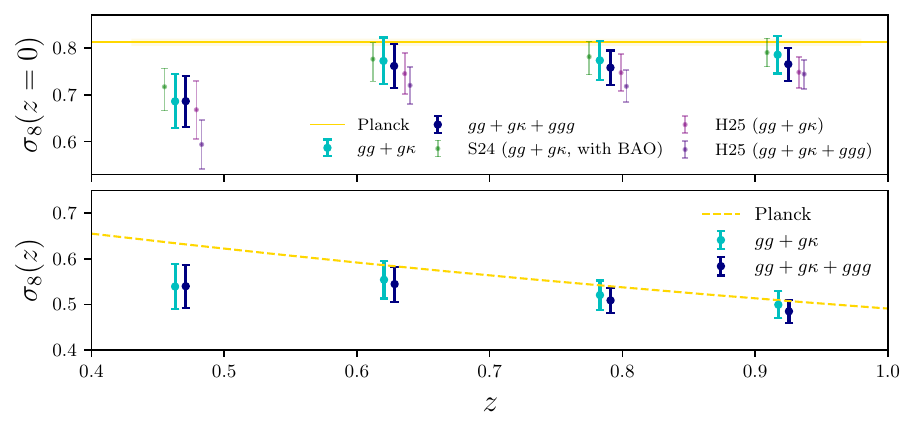}
      \caption{{\bf Tomographic contraints on $\sigma_8$ and growth.} \textit{Top:} constraints on $\sigma_8$ found in each redshift bin using only two-point information and adding the bispectrum measurements (thick points with error bars). These results are compared with those found from the analysis of the same data in S25 \cite{sailer_cosmological_2024} and H25 \cite{harscouet_constraints_2025}, the latter assuming coevolution relations for the bias parameters. \textit{Bottom:} tomographic reconstruction of the growth history in terms of $\sigma_8(z)$. The \planck best-fit prediction is shown in yellow for comparison.}\label{fig:res_singlebins_tomographic}
    \end{figure}
    Figure \ref{fig:res_singlebins_2D} shows the two-dimensional marginalised constraints on $\Omega_m$ and $\sigma_8$ obtained from each individual redshift bin, compared with the constraints found by \planck. Results are shown for two-point functions only and including the \fsb data. We find that the inclusion of three-point information leads to consistently smaller uncertainties by $\sim10$-20\%, in addition to a preference for a smaller value of $\sigma_8$ and larger $\Omega_m$. Interestingly, this preference is also found in full-shape analyses of the three-dimensional clustering of galaxies when including the bispectrum \cite{2112.04515,2206.08327,chudaykin_reanalyzing_2025}. We also find that the data in the first redshift bin favours a value of $\sigma_8$ that is significantly lower than the \planck constraint, by approximately $2.3\sigma$. This is a known result, found in past analyses of this sample by \cite{sailer_cosmological_2024} (S25 hereafter) and \cite{2111.09898}. As reported in H25 \cite{harscouet_constraints_2025}, we find that the galaxy bispectrum measurements are also compatible with this result.

    We compare these tomographic constraints on $\sigma_8$ with previous results in the top panel of Fig. \ref{fig:res_singlebins_tomographic}. Our results are in general good agreement with those found by S25 using the same sample. Our reported uncertainties are $\sim$20\% larger than those of S25 for this tomographic analysis, although the latter are supplemented by external constraints on $\Omega_m$ from baryon acoustic oscillation data. The constraints found by H25 for two-point data are systematically lower than those found here by about $\sim0.5\sigma$, rising to $\sim1\sigma$ when including bispectrum data. These differences are not unexpected, however. First, H25 analysed a previous version of the LRG catalogue compared to that used here and in S25, where a number of shortcomings were addressed (see \cite{zhouDESILuminousRed2023} for further details). Secondly, the constraints obtained by H25 from two-point data modelled the galaxy power spectrum at the tree level, equivalent to a linear bias model. Although this approach may lead to unbiased constraints using sufficiently conservative scale cuts, we saw in Section \ref{ssec:res.sims} that the model is significantly less stable, and past analyses have found consistent downward shifts in amplitude parameters, such as $\sigma_8$, when using a linear bias prescription (e.g. the ``model independent'' analysis of S25). Finally, when including bispectrum data, H25 made use of the coevolution relations of \cite{Lazeyras:2015lgp,Lazeyras:2017hxw} to relate the values of the higher-order bias parameters to $b_1$. Any deviation from these relations would therefore result in additional bias in the inferred cosmological parameters.
    \begin{figure}
    \centering
    \includegraphics[width=0.7\linewidth]{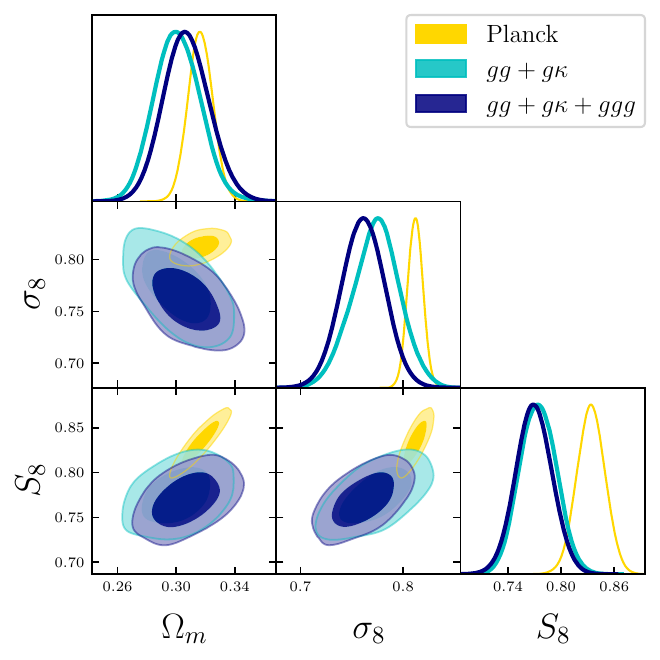}

      \caption{{\bf Combined cosmological constraints.} Constraints on $\sigma_8$, $\Omega_m$, and $S_8$ from the combination of the three highest redshift bins analysed here. Results are shown using only two-point information ($C_\ell^{gg}$ and $C_\ell^{g\kappa}$, \textit{turquoise}), and with the addition of the bispectrum measurements ($\Phi^{ggg}_{LL\ell}$, \textit{dark blue}).}\label{fig:res_bin1_allcomb}
    \end{figure}

    Since, strictly speaking, the clustering measurements made in each redshift bin are only sensitive to the amplitude of matter fluctuations at the redshift of the galaxies that make up the sample, the quantity that is directly measured by our data is $\sigma_8(z)\equiv\sigma_8\,D(z)$, where $D(z)$ is the large-scale linear growth factor\footnote{Note that, in the presence of massive neutrinos, structure growth is scale dependent at the linear level. The scale-dependent features induced by the value of $M_\nu$ assumed here ($M_\nu=0.06\,{\rm eV}$) are negligible given our statistical uncertainties, however.}. We present the tomographic constraints on this parameter obtained from our data in the bottom panel of Fig. \ref{fig:res_singlebins_tomographic}. Our constraints are largely compatible with the growth history predicted by \planck (also shown in the figure), with the exception of the mild $\sim2\sigma$ downwards deviation observed in the first redshift bins. This is also in agreement with other constraints on the growth history from CMB lensing tomography, including S25 (using the same galaxy sample), \cite{2105.12108} using a variety of galaxy samples, \cite{2309.05659,2409.02109} using the unWISE catalog, \cite{2402.05761,2507.08798} using high-redshift quasars from {\sl Quaia}, \cite{2503.24385} combining photometric LRGs and the DESI Bright Galaxy Sample and, most recently, \cite{2510.09563} using the 2MPZ and WISE$\times$SuperCOSMOS galaxy samples at low redshifts. The addition of the galaxy bispectrum results in a $\sim10$-15\% reduction in the statistical uncertainties of these measurements, particularly at the higher redshifts, accompanied by the slight downwards shift in the value of $\sigma_8(z)$ discussed earlier.

\begin{figure}
      \centering
      \includegraphics[width=\textwidth]{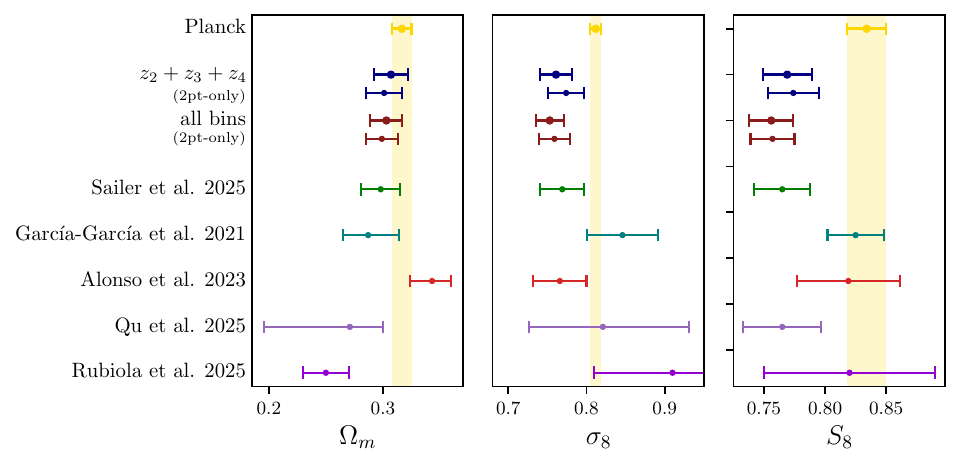}
      \caption{{\bf Comparison with past constraints.} Comparison of our constraints on $\{\Omega_m, \, \sigma_8,\,S_8\}$ with past constraints derived from lensing tomography \cite{sailer_cosmological_2024, 2105.12108, 2306.17748, 2410.10808, 2510.09563} as well as the \planck results.}\label{fig:whisker_plot}
    \end{figure}

    Finally, we combine the measurements in different redshift bins to place joint constraints on $\sigma_8$ and $\Omega_m$. Given the mild tension between the constraints found in the first redshift bin and the rest, we choose our fiducial constraints to correspond to the combination of the last three redshift bins, although we will also present the results of combining all redshift bins. Fig. \ref{fig:res_bin1_allcomb} shows these combined constraints, including those found using only two-point correlations, as well as those incorporating the galaxy bispectrum. Additionally, Fig. \ref{fig:whisker_plot} shows the same constraints, as well as those arising from the combination of all redshift bins, in comparison with other results found in the literature. Our constraint on $\Omega_m$ is compatible with \planck within $\sim1\sigma$. As shown in \cite{harscouet_constraints_2025,2306.17748,2510.09563}, this constraint is dominated by the large-scale broadband shape of the power spectrum, as constrained by $C_\ell^{gg}$ and $C_\ell^{g\kappa}$. Perhaps interestingly, our constraints on this parameter lie on the low side of the \planck measurements, as do most estimates derived from CMB lensing tomography at redshifts $z\lesssim1$, as well as the latest BAO constraints on $\Lambda$CDM from DESI \cite{2503.14738}. 
    Nevertheless, the measurement uncertainties are too large to draw any definitive conclusion regarding a potential parameter tension.

    In turn, our fiducial constraints on $\sigma_8$ are lower than \planck at the $2.4\sigma$ level when including power spectrum and bispectrum measurements. Compared to the $1.6\sigma$ deviation obtained using only two-point functions, this result is driven by the $\sim 15\%$ smaller error bars combined with the $\sim0.5\sigma$ downward shift in $\sigma_8$ we discussed above. Combining the measurements from all redshift bins, this tension in the value of $\sigma_8$ rises to the $3\sigma$ level ($2.5\sigma$ from two-point functions only), driven by the significantly lower amplitude of matter fluctuations favoured by the first redshift bin.
    
    We can also derive constraints on the so-called ``clumping parameter'' $S_8\equiv\sigma_8\sqrt{\Omega_m/0.3}$, which can be accurately constrained by cosmic shear analyses. Combining different redshift bins, we obtain:
    \begin{equation}
      S_8=0.769 \pm 0.020\,\,(z_2+z_3+z_4),\hspace{12pt}
      S_8=0.756 \pm 0.018\,({\rm all\,\,bins}).      
    \end{equation}
    Compared to the \planck constraints on this parameter ($S_8=0.832\pm0.013$), these measurements lie 2.6 and 3.4$\sigma$ lower, respectively. The tension in the latter case is driven by the inclusion of the first redshift bin, where the constraints on both $\Omega_m$ and $\sigma_8$ lie below the \planck values (see top left panel of Fig. \ref{fig:res_singlebins_2D}). This level of tension is marginally higher than that found in \cite{sailer_cosmological_2024} using lensing from \planck and ACT ($2.3\sigma$), driven primarily by the shift induced by the addition of bispectrum data, as well as the different modelling choices. 
    Fig. \ref{fig:whisker_plot} shows our constraints on $\Omega_m$ and $\sigma_8$ in the context of other CMB lensing tomography analyses, including S25 \cite{sailer_cosmological_2024}, the analysis of \cite{2105.12108} combining multiple galaxy samples, the constraints of \cite{2410.10808} using an alternative sample from the Legacy Survey, the recent analysis of \cite{2510.09563} using low-redshift data, and the high-redshift constraints of \cite{2306.17748} using quasars from {\sl Quaia}. The constraints found here are in broad agreement with these works, although it is interesting to note that the analyses showing the strongest tension in $\sigma_8/S_8$ with the CMB are those using LRG samples. Additionally, with the exception of \cite{2306.17748}, most lensing tomography analyses, in the absence of external BAO data, seem to favour a marginally low value of $\Omega_m$. Given the current uncertainties, this could be simply a mild statistical anomaly, although misspecification in the non-linear bias model, or the presence of systematic contamination in the observed fluctuations in galaxy number counts, could also contribute to a low $S_8$ or $\Omega_m$.

    To further explore the limits of the bias model used here, and the ability of the bispectrum measurements to improve the final cosmological constraints, we have explored the possibility of using simpler bias models to describe the data. In particular, we considered two cases: one in which the counter-term parameters $\{c_m,\,b_{k^2}\}$ and the third-order bias parameter $b_{3nl}$ are set to zero, and one in which the higher-order bias parameters $\{b_2,\,b_s,\,b_{3nl}\}$ are set to zero. In the first case, the power spectra and bispectra are both described by the same set of bias parameters, thus maximising the information content in the bispectrum. The details of this study are discussed in Appendix \ref{app:tests.simplebias}, where we show that these simplified models are able to describe the data well over the same range of scales as our fiducial bias model. When applied to the real data, we (unsurprisingly) recover tighter constraints on both $\Omega_m$ and $\sigma_8$, with a small shift towards the values preferred by \planck. Interestingly, as shown in Table \ref{tab:results_table}, both models (indicated by the subset of parameters that we vary) are able to fit the data well, with a very small penalty in the best-fit $\chi^2$ with respect to the fiducial model, commensurate with the reduced number of degrees of freedom. This suggests the additional complexity of the fiducial model is not required by the data. Nevertheless, a more detailed validation of these simpler models, and their ability to describe the clustering of generic galaxy samples, would be required before they can be used reliably in cosmological analyses.

  \subsection{Galaxy bias relations from projected clustering}\label{ssec:res.bias}
    Galaxy bias is a fundamental element of perturbation theory models for non-linear clustering. In this first-principles approach, the bias coefficients have to be treated as nuisance parameters in cosmological inference, introducing degeneracies that inevitably limit the precision of parameter estimation. Moreover, non-linear bias parameters notably contribute to the power spectrum in a largely degenerate way, making it challenging to independently constrain them using power spectrum alone in 3D analyses. This problem becomes more pronounced for projected clustering, where linear features such as the BAO wiggles are highly suppressed. In this context, the inclusion of higher-order correlation functions helps mitigate this limitation: the need to simultaneously describe two- and three-point functions helps break the degeneracies between different bias parameters, allowing us to measure them at higher precision. A better constraint on the higher-order bias parameters may then allow us to use the galaxy-galaxy and galaxy-$\kappa$ power spectra on mildly non-linear scales to constrain cosmology. 
    \begin{figure}
      \centering
      \begin{subfigure}[t]{0.48\textwidth}
        \centering
        \includegraphics[width=\linewidth]{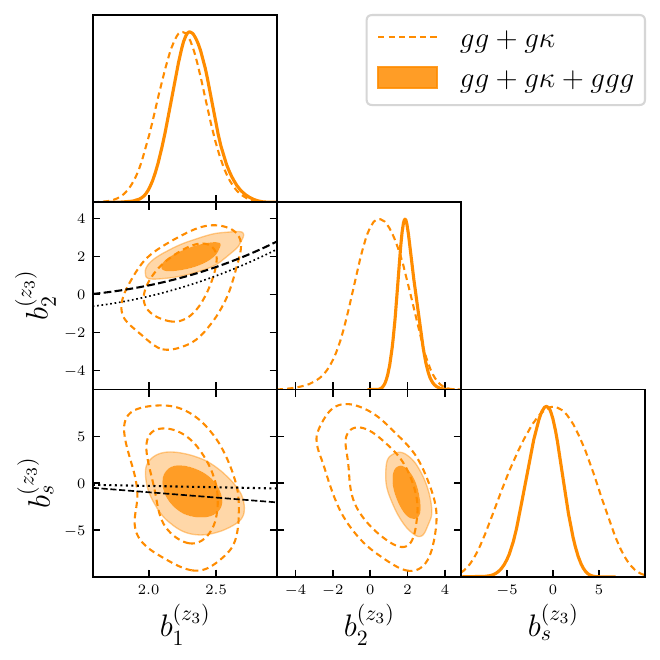}
      \end{subfigure}
      \hfill
      \begin{subfigure}[t]{0.48\textwidth}
        \centering
        \includegraphics[width=\linewidth]{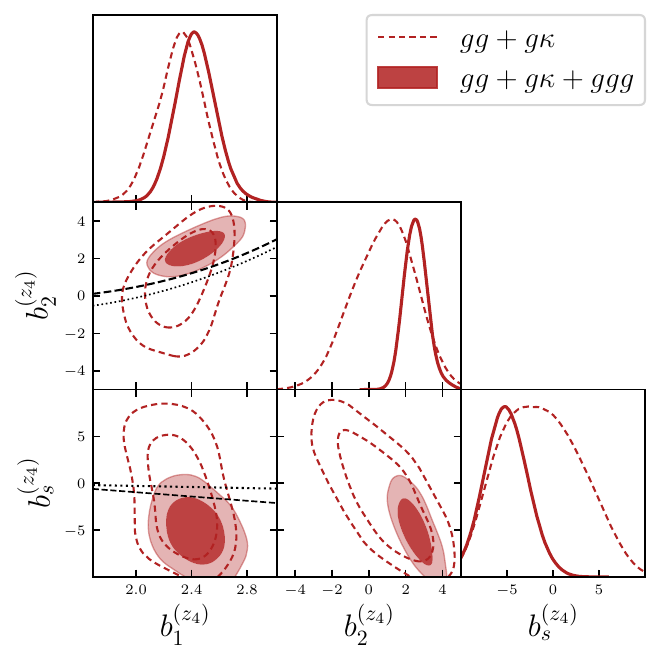}
      \end{subfigure}
      \caption{{\bf Constraints on bias parameters.} Impact of additional three-point information on the posterior of the linear and higher-order bias parameters bias parameters $\{b_1,\,b_2,\,b_s\}$ that contribute to our model of the bispectrum (Eq. \ref{eq:Bggg-theory}). Results are shown for the two highest redshift bins, with similar results found for the first two bins. The inclusion of three-point information leads to a factor $\mathcal{O}(5-10)$ improvement in the FoM for this set of bias parameters. The coevolution relations obtained from \cite{Lazeyras:2015lgp,Lazeyras:2017hxw} for haloes are shown as dotted lines, while the relations found by \cite{barreira_galaxy_2021} for galaxies are shown as dashed lines.}\label{fig:bias_posterior}
    \end{figure}

    Fig. \ref{fig:bias_posterior} shows the constraints on the bias parameters using only power spectra or power spectra and bispectrum measurements. Results are shown for the last two redshift bins, where we are able to use a larger number of angular modes at a fixed $k_{\rm max}^B$ and where, in consequence, this improvement is more evident. Quantitatively, the inclusion of bispectrum information, even on the conservatively large scales used here, leads to a remarkable $\mathcal{O}(10)$ improvement in the FoM for the three bias parameters $\{b_1,b_2,b_s\}$, with $b_2$ and $b_s$ experiencing the strongest reduction in their statistical uncertainties. Importantly, the constraints obtained from the full data vector are compatible with those obtained from power spectra alone, with the bispectrum favouring the region of parameter space with larger values of $b_1$ and $b_2$ (and marginally lower $b_s$). This trend leads to the mild downwards shift in $\sigma_8$ observed in the previous section when including bispectrum data (as also observed in \cite{chudaykin_reanalyzing_2025}). Also, and reassuringly, the best-fit bias parameters $\{b_1,b_2,b_s\}$ found using power spectrum data alone provide a reasonable prediction for the bispectrum measurements (blue lines in Fig. \ref{fig:fit-and-chi2}), demonstrating the internal self-consistency of our data and model.
    \begin{figure}
      \centering
      \includegraphics[width=\textwidth]{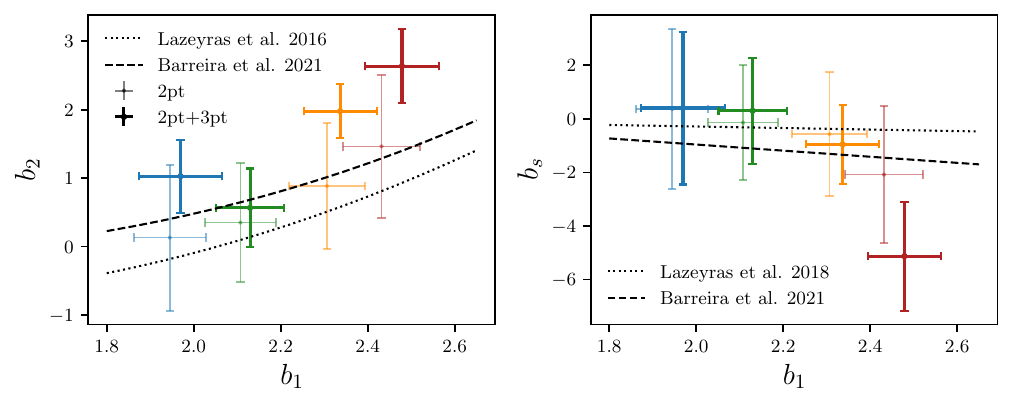}
      \caption{{\bf Bias relations.} Relation between the quadratic ($b_2$, \textit{left}) and tidal ($b_s$, \textit{right}) bias parameters and the linear bias $b_1$ found in each redshift bin. Results are shown using only two-point information (\textit{thinner, shaded markers}), and with the inclusion of bispectrum data (\textit{thicker markers}). These constraints are compared against the coevolution relations for haloes (\cite{Lazeyras:2015lgp,Lazeyras:2017hxw}, \textit{dotted lines}), and galaxies (\cite{barreira_galaxy_2021}, \textit{dashed lines}).}\label{fig:coevolution}
    \end{figure}
    Having improved constraints allows us to examine the relationships between the different bias parameters. In particular, measurements of the clustering of galaxies and haloes in $N$-body simulations have found that the higher-order bias parameters exhibit a correlation with the linear bias $b_1$. These often called ``coevolution relations'' are not completely universal, differing slightly between haloes and galaxies, and showing a non-negligible scatter. Nevertheless, it is interesting to confirm whether the clustering of galaxies observed in real data also obeys these relations. Fig. \ref{fig:bias_posterior} shows the relations found by \cite{Lazeyras:2015lgp} and \cite{Lazeyras:2017hxw} for haloes (dotted lines), and by \cite{barreira_galaxy_2021} for galaxies (dashed lines). The preference of the bispectrum data for a larger $b_2$ pushes our constraints marginally away from the coevolution relations. This is shown in more detail in Fig. \ref{fig:coevolution}, which collects the constraints found in all redshift bins when analysed jointly. 
    With the inclusion of bispectrum data, our constraints show a mild preference (at the $\sim2\sigma$ level) for upwards (downwards) deviations in $b_2$ ($b_s$) for larger values of $b_1$. These deviations are not entirely unexpected, given the relatively large scatter found in the case of galaxies \cite{barreira_galaxy_2021,2110.05408,2412.01888}. Note also that our constraints use data from a single galaxy type (LRGs), spanning a relatively narrow range of $b_1$ values.

\section{Conclusions}\label{sec:conc}
  We have carried out the first comprehensive cosmological analysis of lensing tomography incorporating information from the angular galaxy bispectrum, complementing the first tentative analysis of \cite{harscouet_constraints_2025}. A key aspect of such an analysis is the high redundancy of the full data vector, forcing our model of the galaxy-matter connection to simultaneously describe the two- and three-point clustering of galaxies as well as their cross-correlation with the matter overdensities. This redundancy improves both the precision and robustness of the resulting cosmological constraints.

  We have used a perturbative bias expansion within the Effective Field Theory of LSS framework, modelling the galaxy-galaxy and galaxy-convergence power spectra at one-loop order, and the galaxy bispectrum at tree level. We have validated this model against $N$-body-based mock observations, and determined the scales over which it can safely recover cosmological constraints. Compared to studies of the three-dimensional clustering of galaxies, this task is made simpler in the case of projected statistics due to the negligible impact of non-linear redshift-space distortions. We have shown that the model is able to recover unbiased constraints down to scales where cosmological information is in fact saturated for the LRG sample studied here.

  Applying this model to the analysis of the four different redshift bins considered here, we find constraints on $\sigma_8$ and $\Omega_m$ that are consistent with past works analysing this sample, and have shown that the inclusion of the galaxy bispectrum leads to a $\sim10$-15\% improvement on the cosmological constraints. The constraints on $\Omega_m$ and $\sigma_8$ found from the three highest-redshift bins are compatible with each other and in reasonable agreement with \planck, as is the growth history reconstructed from these data (see Fig. \ref{fig:res_singlebins_tomographic}). In turn, the first redshift bin shows a preference for lower $\sigma_8$, and is in tension with \planck at the $\sim2.5\sigma$ level.

  We have then derived cosmological constraints from the joint analysis of the last three redshift bins. The measured value of $\sigma_8$ ($S_8$) is moderately lower than the \planck value at the $2.4\sigma$ ($2.6\sigma$) level. This is due to the mild but coherent preference for a lower $\sigma_8$ displayed by the two-point clustering of galaxies in all redshift bins, combined with the additional $\sim0.5\sigma$ downwards shift observed when including the galaxy bispectrum. Interestingly, a similar shift has been observed in the three-dimensional clustering of galaxies when combining power spectrum and bispectrum data \cite{chudaykin_reanalyzing_2025}. This tension is then exacerbated, above the $3\sigma$ level, when combined with the first redshift bin. In agreement with the results obtained in our per-bin analysis, the inclusion of the projected bispectrum leads to a $\sim10\%$ improvement in the statistical uncertainties on cosmological parameters.

  Including the bispectrum has a much stronger impact on the higher-order bias parameters of our model, $b_2$ and $b_s$, with a reduction in their statistical uncertainties of more than a factor 2 in some cases. This allows us to study the relations between the different bias parameters with higher precision, as well as their agreement with coevolution relations found in simulations. The constraints obtained after including bispectrum information are in good agreement with those from power spectra alone, but show a consistent preference for higher $b_1$ and $b_2$, and lower $b_s$. This manifests itself as a mild departure from the coevolution relations of \cite{barreira_galaxy_2021} for high values of $b_1$. This is not entirely surprising, however, given the large scatter expected around these relations.

  Adding bispectrum data improves the robustness and precision of the resulting cosmological constraints. The improvement in the latter case is relatively disappointing, however ($\sim10$-15\%), and it is interesting to reflect on the potential reasons for this. First, although the bispectrum is able to significantly improve the constraints on $b_2$ and $b_s$ when modelled at the tree level, the counter-term parameters $\{b_{k^2},\,c_m\}$ and the third-order parameter $b_{3nl}$ are only constrained by the power spectrum measurements. This leaves significant freedom to describe the featureless clustering of galaxies on mildly non-linear scales. In Appendix \ref{app:tests.simplebias} we have explored the possibility of using a simpler bias model, setting $\{b_{k^2},\,c_m,\,b_{3nl}\}$ to zero, and thus allowing us to maximally exploit the information contained in the bispectrum measurements. Interestingly, we have shown that this simplified model is still able to recover unbiased constraints in mock data and, when applied to the data, the aforementioned tensions are somewhat relieved. Nevertheless, the relative improvement in the final constraints from including the bispectrum is still limited to $\sim10\%$, suggesting that the limiting factor is in fact not the additional freedom allowed by the counterterm parameters and $b_{3nl}$.
  
  Another potential cause for the mild improvement from the bispectrum is the large uncertainties in $C_\ell^{g\kappa}$ and $\Phi^{ggg}_{LL\ell}$ relative to the galaxy power spectrum. Both of these measurements are crucial to break degeneracies in the bias model and constrain cosmology, and the final constraints may be limited by these uncertainties, rather than freedom in the model itself. Future work could improve both of these aspects of our analysis using existing data. First, replacing the CMB lensing convergence with cosmic shear data from existing imaging surveys \cite{2011.03408,2503.19442} could improve the sensitivity of the galaxy-lensing power spectrum. It could also allow us to exploit the galaxy-galaxy-lensing bispectrum, which we have discarded in this analysis due to the low signal-to-noise ratio on the large scales over which it can be modelled accurately, as shown in \cite{harscouet_constraints_2025}. Secondly, our measurements of the angular bispectrum have been based on the \fsb estimator presented in \cite{harscouet_fast_2025}, which only recovers triangle configurations that are close to isosceles. As discussed in Section \ref{ssec:data.cl} and \cite{harscouet_fast_2025}, the \fsb estimator can be easily generalised to recover the remaining non-zero bispectrum configurations, thus increasing the total three-point statistical power. Finally, our use of the bispectrum is limited by the range of scales over which our tree-level model is sufficiently accurate. In the future, this could be significantly improved by resorting to more sophisticated approaches to galaxy bias, such as hybrid effective field theory (HEFT, \cite{0807.1733,1910.07097,2101.12187}), which could remain accurate on significantly smaller scales without a substantial increase in the number of bias parameters required.

  This work demonstrates that a joint analysis of the projected power spectrum and bispectrum of galaxies and weak lensing can be efficiently and reliably carried out with existing data. The deployment of these methods could also enhance the science returns of future imaging surveys, such as the Rubin Observatory \cite{0805.2366,1809.01669}, maximising the information extracted from the projected clustering of galaxies.

\acknowledgments
  We thank Carlos Garc\'ia-Garc\'ia, Andrina Nicola, Sergi Novell-Massot, Noah Sailer, Emiliano Sefusatti and An\v{z}e Slosar for useful comments and discussions.  LH is supported by a Hintze studentship, which is funded through the Hintze Family Charitable Foundation. MZ and DA acknowledge support from the Beecroft trust. We made extensive use of computational resources at the University of Oxford Department of Physics, funded by the John Fell Oxford University Press Research Fund.

\appendix
  \section{Additional tests}\label{app:tests}

  \subsection{Treatment of Redshift-space distortions}\label{app:tests.rsd}
\begin{figure}
      \centering
      \includegraphics[width=\linewidth]{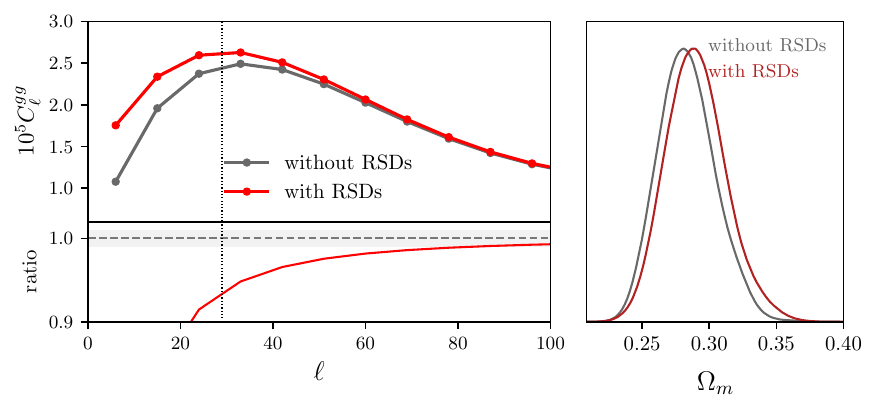}
      \caption{{\bf The impact of RSDs.} \textit{Left:} contribution of RSDs to the galaxy power spectrum on large scales. The vertical dotted line represents the large-scale cut that we adopt in our fiducial analysis. \textit{Right:} systematic shift on the inferred $\Omega_m$ parameter for the LRGs $z_3$ bin only when ignoring the RSD contribution. Using our fiducial large-scale cut, the shift is relatively minor ($\sim0.3\sigma$).}\label{fig:RSDs-impact}
    \end{figure}

    Redshift-space distortions are the most relevant secondary effect modifying the galaxy angular power spectrum on large scales, particularly for relatively narrow redshift bins \cite{1812.05995,tanidis_developing_2019}. As described in Section \ref{sssec:meth.theory.cl}, we include RSDs at the linear level in the galaxy auto-spectrum, including only the correlations between the RSD term and the linear density term proportional to $b_1$ and the RSD auto-correlation. Given the results of our simulation-based validation, this is sufficient to obtain unbiased cosmological constraints from our data. Nevertheless, to gain a better intuitive understanding of the impact of RSDs on our analysis, the left panel of Fig. \ref{fig:RSDs-impact} shows the large-scale power spectrum with and without the RSD contributions. Within the scale cuts used here, RSDs lead to deviations in the power spectrum amplitude of up to $\sim10\%$, peaking on the largest scales used and becoming negligible at high $\ell$.
    
    Repeating our analysis omitting these contributions, the right panel of Fig. \ref{fig:RSDs-impact} shows that the largest effect is a relatively mild ($\sim0.3\sigma$) downwards shift in the preferred value of $\Omega_m$. Thus, although relevant for this analysis, a precise modelling of RSDs (e.g. including the impact of non-linear ``fingers of God'') is not required.

  \subsection{Simpler bias parametrisations}\label{app:tests.simplebias}
    \begin{figure}
      \centering
      \begin{subfigure}[t]{0.47\textwidth}
        \centering
        \includegraphics[width=\linewidth]{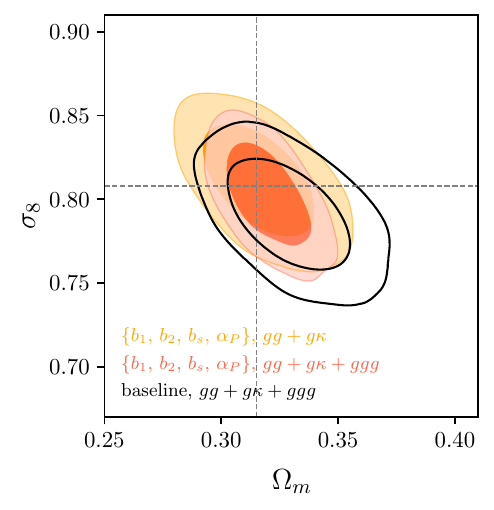}
      \end{subfigure}
      \begin{subfigure}[t]{0.47\textwidth}
        \centering
        \includegraphics[width=\linewidth]{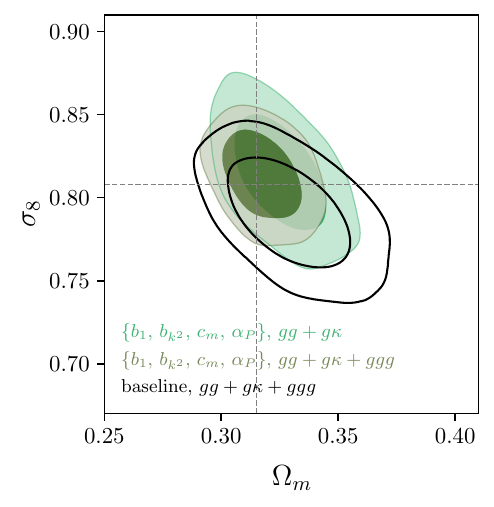}
      \end{subfigure}
      \caption{{\bf Simpler bias models on simulations.} Constraints on $\Omega_m$ and $\sigma_8$ obtained from the \absum simulation, combining all redshift bins. Our baseline constraints are shown as black lines. The brighter and darker contours show the constraints found using only two-point functions and combining them with the bispectrum, respectively. The \textit{left panel} shows the constraints found using a simpler bias model in which the scale-dependent counterterms $\{b_{k^2},\,c_m\}$, and the third-order parameter $b_{3nl}$ are set to zero. The right panel shows the constraints found setting $b_2$, $b_s$, and $b_{3nl}$ to zero instead, in a model inspired by the phenomenological parametrisation of \cite{2508.05319}. Both restricted parametrizations are able to find unbiased constraints on the cosmological parameters.}\label{fig:restricted_sims}
    \end{figure}
    \begin{figure}
      \centering
      \begin{subfigure}[t]{0.46\textwidth}
        \centering
        \includegraphics[width=\linewidth]{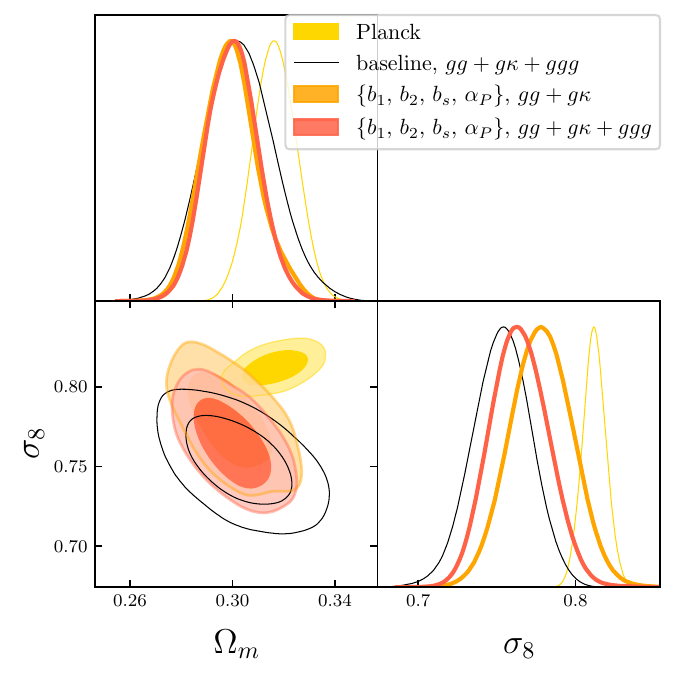}
      \end{subfigure}
      \begin{subfigure}[t]{0.46\textwidth}
        \centering
        \includegraphics[width=\linewidth]{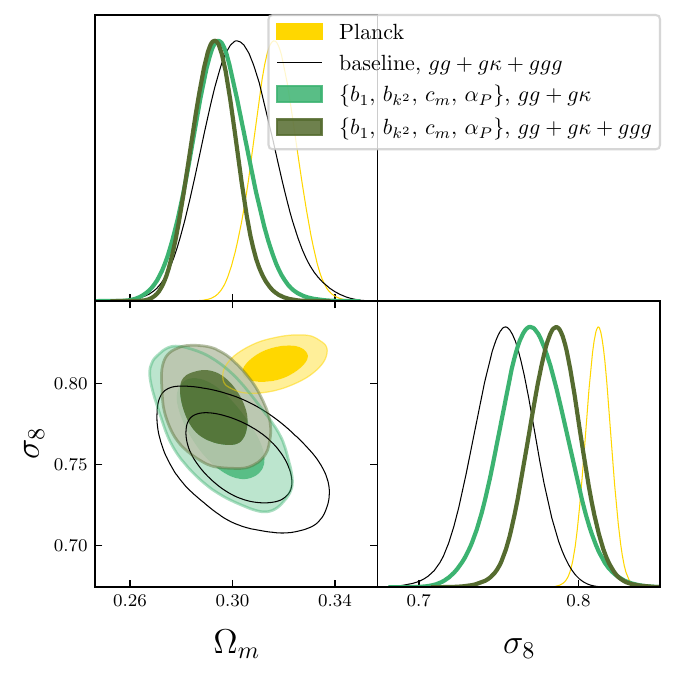}
      \end{subfigure}
      \caption{{\bf Simpler bias models on real data.} As Fig. \ref{fig:restricted_sims}, now showing the constraints found when analysing the real data. The \planck constraints are also shown for comparison.}\label{fig:restricted_realdata}
    \end{figure}
    The bias model used here (see Section \ref{sssec:meth.theory.bias}), in which power spectra are modelled at the 1-loop level including EFT counterterms, and the bispectrum is modelled at the tree level, involves free parameters, such as $c_m$, $b_{k^2}$, and $b_{3nl}$, which can only be constrained by a single type of correlation (e.g. two-point functions, but not the bispectrum). Although this level of freedom is theoretically necessary and physically motivated, it is interesting to explore whether simpler versions of this model are still able to provide unbiased cosmological constraints. Furthermore, recent works \cite{2508.05319} have shown that, in the case of real-space galaxy clustering, a simplified phenomenological parametrization of galaxy bias is still able to provide an adequate and unbiased description of the data. In this section we explore how incorporating bispectrum information impacts parameter constraints when working within two restricted bias parametrizations.

    Specifically, we test two complementary simplified models. First, we explore a parametrization that only retains the parameters that enter in Eq. \eqref{eq:Bggg-theory}, i.e. varying $\{b_1,\,b_2,\, b_s,\, \alpha_P\}$, while fixing the counterterm values, and third order Eulerian bias, to zero. Then, we also tested a theoretical model that, in the spirit of \cite{2508.05319}, effectively only parametrizes the deviations from a scale-independent linear bias through counterterm-like coefficients. In our setup, we achieve this by only varying the subset of nuisance parameters $\{b_{1},\,b_{k^2},\,c_m,\,\alpha_P\}$, fixing to zero the remaining bias parameters\footnote{Note that the model in \cite{2508.05319} also allows for a stochastic noise component in the galaxy-matter power spectrum, making it more general than the model studied here.}.

    We begin by testing its performance on simulated data. As shown in Fig. \ref{fig:restricted_sims}, we find that both simplified parametrizations are able to recover the fiducial values of the cosmological parameters when applied to the combination of all redshift bins, both using only two-point correlators and including the bispectrum. This is likely due to the relatively featureless nature of the projected galaxy statistics, making it possible to describe them with a relatively small number of smooth ``basis'' functions. Notably, in this simulated data, the addition of $ggg$ information is perfectly compatible with the 2-point posterior, tightening the constraints around the $(\Omega_m,\, \sigma_8)$ fiducial point, improving their statistical uncertainties by $(26\%,\,8\%)$ using the $\{b_1,b2,b_s,\alpha_P\}$ model (left panel) and $(17\%,\,24\%)$ using the $\{b_1,b_k^2,c_m,\alpha_P\}$ model (right panel).

    Then, having validated both on simulations, we apply them to the analysis of the real clustering data. Fig. \ref{fig:restricted_realdata} shows the posterior distribution of the cosmological parameters, as well as the constraints found using our baseline model for reference. We find that, using the $\{b_1,\,b_2,\,b_s,\,\alpha_P\}$, the uncertainty on $\Omega_m$ shrinks by $\sim25\%$ with respect to our baseline model, while the $\sigma_8$ remains largely unchanged. The value also favours a slightly lower value of $\Omega_m$ and a larger value of $\sigma_8$, reducing the tension with \planck found in our baseline analysis. A similar shift is found in the case of the $\{b_1,\,b_{k^2},\,c_m,\,\alpha_P\}$ model, again relieving this tension. The inclusion of the bispectrum in this model also leads to a significant $\sim20\%$ reduction in the $\sigma_8$ uncertainty.

    It is worth stressing the fact that these two simpler bias models, defined as subsets of the full EFT bias model presented, are not, strictly speaking, well motivated or complete from a physical point of view. Thus, their ability to describe our data, as tested on simulations, might be purely accidental and only applicable to the specific galaxy sample we have simulated, which would lead to artificially reduced cosmological constraints. Therefore, a more comprehensive validation of the models, considering a large suite of non-linear galaxy clustering scenarios, would be necessary before they can be safely applied to cosmological analyses.

\subsection{Emulators accuracy}\label{ssec:em.acc}
The approach of emulating, rather than computing \textit{ab initio} at each likelihood step, the theory templates enables fast and feasible MCMC sampling as detailed in Sect. \ref{ssec:meth.emu}. While essential, this additional step may introduce systematic errors; therefore, the accuracy of the emulators must be carefully validated. 

\begin{figure}
    \centering
\includegraphics[width=\linewidth]{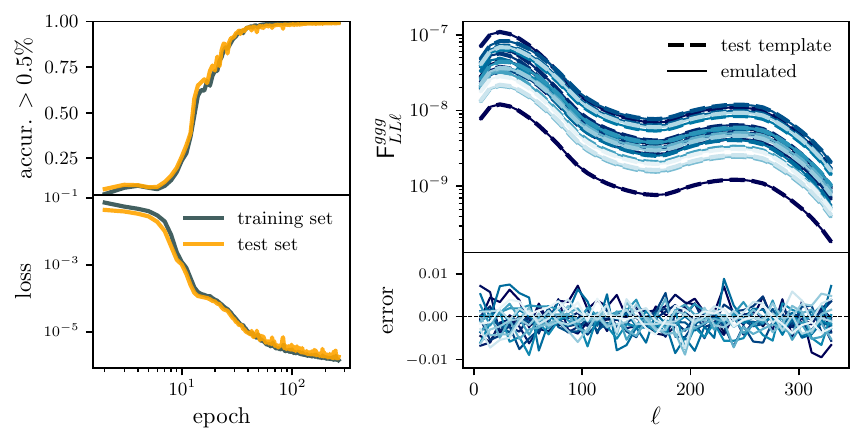}
\caption{{\bf Emulating \fsb templates.} Training history (\textit{left}) of the emulator for the $\mathsf{F}^{ggg}_{LL\ell}$ (first filter) template, and performance (\textit{right}) on some elements of the test set. All the emulators for the other templates present an analogue training history and a similar performance.
 }\label{fig:fsb-emu}
\end{figure}

To this purpose, we focus on the emulator of the $\mathsf{F}^{ggg}_{LL\ell}$ template as a representative example, as it exhibits consistent behaviour with the others.
Specifically, in Fig. \ref{fig:fsb-emu} we show the evolution of the loss function (mean squared errors) and of our accuracy metric throughout the training epochs. We find that in a few hundred epochs the loss function improves by almost 5 orders of magnitude, and virtually 100\% of the output layer neurons differ less than 0.5\% from the ground truth (which prompts us to stop the training). We repeat this exercise for both the training set and the test set (which is never seen by the Neural Network while training), finding similar results and no indication of overfitting. 

Finally, in the right panel of Fig. \ref{fig:fsb-emu}, we show the overall accuracy of this emulator, in terms of percentage differences between a subsample of the $\mathsf{F}^{ggg}_{LL\ell}$ template values used in the test set and their corresponding emulator predictions. We find that the emulator reproduces the test set at better than 1\% over the whole considered $\ell$ range, with no significant scale dependence. We find similar performance for the emulators of all of the templates of section \ref{ssec:meth.emu}, which is accurate enough given the statistical error of the dataset used in this work.

\bibliography{biblio}{}

\end{document}